\documentclass[a4paper,fleqn,usenatbib,useAMS]{mnras}

\usepackage{graphicx}	
\usepackage{amsmath}	
\usepackage{amssymb}	
\usepackage{multicol}        
\usepackage{bm}		
\usepackage{pdflscape}	
\usepackage{todonotes}

\usepackage[T1]{fontenc}
\usepackage{ae,aecompl}

\usepackage{newtxtext,newtxmath}

\title[KHI and cold accreting filaments]{
The density distribution of accreting cosmic filaments as shaped by Kelvin-Helmholtz instability
}

\author[A. C. E. Vossberg et. al.]{Ann-Christine E. Vossberg$^{1}$
\thanks{Contact e-mail: \href{mailto:ann-christine.vossberg@phys.ethz.ch}{ann-christine.vossberg@phys.ethz.ch}}
, Sebastiano Cantalupo $^{1}$, 
Gabriele Pezzulli $^{1}$
\\
$^{1}$Department of Physics, ETH Zurich, Wolfgang-Pauli-Strasse 27, 8093 Zurich, Switzerland}

\date{Last updated 2019 April 3; in original form 2019 April 3}

\pubyear{2018}

\begin{document}
\label{firstpage}
\pagerange{\pageref{firstpage}--\pageref{lastpage}}
\maketitle

\begin{abstract}
Cosmic filaments play a crucial role in galaxy evolution transporting gas from the intergalactic medium into galaxies. However, little is known about the efficiency of this process and whether the gas is accreted in a homogenous or clumpy way.
Recent observations suggest the presence of broad gas density distributions in the circumgalactic medium which could be related to the accretion of filaments. 
By means of high-resolution hydrodynamical simulations, we explore here the evolution of cold accreting filaments flowing through the hot circumgalactic medium (CGM) of high-z galaxies.
In particular, we examine the nonlinear effects of Kelvin-Helmholtz instability (KHI) on the development of broad gas density distributions and on the formation of cold, dense clumps. 
We explore a large parameter space in filament and perturbation properties, such as, filament Mach number, initial perturbation wavelength, and thickness of the interface between the filament and the halo.
%
We find that the time averaged density distribution of the cold gas is qualitatively consistent with a skewed log-normal probability distribution function (PDF) plus an additional component in form of a high density tail for high Mach-numbers.  
Our results suggest a tight correlation between the accreting velocity and the maximum densities developing in the filament which is consistent 
with the variance-Mach number relation for turbulence. Therefore, cosmological accretion could be a viable mechanism to produce turbulence and broad gas density distributions within the CGM.

\end{abstract}




\begingroup
\let\clearpage\relax
\endgroup
\newpage

\section{Introduction}

Our standard cosmological $\Lambda$CDM model predicts that both dark and baryonic matter is arranged into a system of filaments, called the ``cosmic web" \citep[e.g.][]{Bond1996}. Recent observations have found extended Ly$\alpha$ emission around quasars at redshifts $z \sim 3$ that is believed to trace the densest parts of this cosmic web \citep[e.g.][]{Cantalupo2014, Martin2014a, Hennawi2015, Borisova2016}. 
Clustering studies \citep[e.g.][]{Eftekharzadeh2015} suggest that these quasars are hosted by haloes massive enough to be 
above the critical mass for shock heating \citep{Birnboim2003, Dekel2006} and therefore containing mostly hot gas at temperatures of $T>10^6$ K. 
Theoretical arguments and cosmological simulations \citep{Keres2005, Ocvirk2008, Dekel2009, Ceverino2010, Faucher2011, Voort2011} suggest the presence,
within these hot haloes, of cold accreting filaments possibly feeding the central galaxy and driving its evolution. 
However, it is unclear what happens to the cold accreting filaments along their way through the hot CGM. 
Ly$\alpha$ and metal emission observations of the CGM at $z>2$ suggest that the cold gas (T$\simeq10^4$ K) has a broad density distribution with densities greatly exceeding $n>1 \; \rm{cm}^{-3}$ \citep{Cantalupo2014, ArrigoniBattaia2015, Borisova2016, Hennawi2015, Cai2017, Cantalupo2018}.
What is the origin of such broad gas density distribution within the CGM? Is cosmological accretion of cold filaments responsible for the presence of dense, cold clumps in the CGM?
If properly resolved, accreting filaments might not stay laminar, as typically found in cosmological simulations, but could be subject to hydrodynamic and other instabilities that would result in the development of turbulence and fragmentation. 
In particular, Kelvin-Helmholtz instability (KHI) is expected to occur at the interface between the accreting filament and the hot halo \citep[e.g.][]{Bodo1998}.  
Recently, a series of studies have focused on the effects of KHI on accreting filaments. These authors have developed a thorough analytical and numerical model on the stability of cold accreting filaments \citep{Mandelker2016, Padnos2018, Mandelker2018}.  
They have found that in some conditions KHI is expected to become highly nonlinear before the filament reaches the central galaxy. Filaments with a size in the range of 0.5 - 5 \% of the virial halo radius fully disrupt prior to reaching the central galaxy, leading to powerful turbulence around the stream. 

In this paper we study the nonlinear evolution of KHI in cold accreting filaments, using idealised simulations, for a purely adiabatic case. We complement \cite{Mandelker2016}'s and \cite{Padnos2018}'s work by studying the density of the filament, particularly focusing on the high densities, developing due to KHI. 

In Section \ref{sec_methods} we will describe the methods and initial setup used for the simulations. We present our results in Section \ref{sec_results} followed by a discussion in Section \ref{sec_discussion}. Lastly, we summarise our findings in Section \ref{sec_summary}.

\section{Methods}
\label{sec_methods}

\begin{figure}
  \includegraphics[width=\columnwidth]{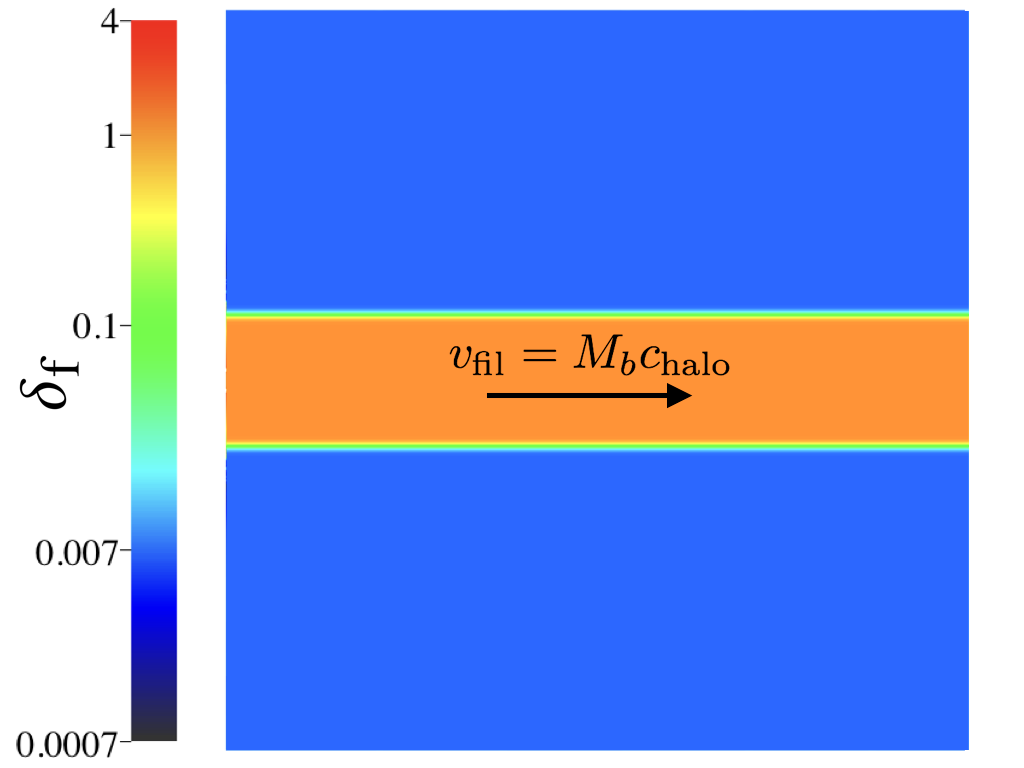}
   \includegraphics[width=\columnwidth]{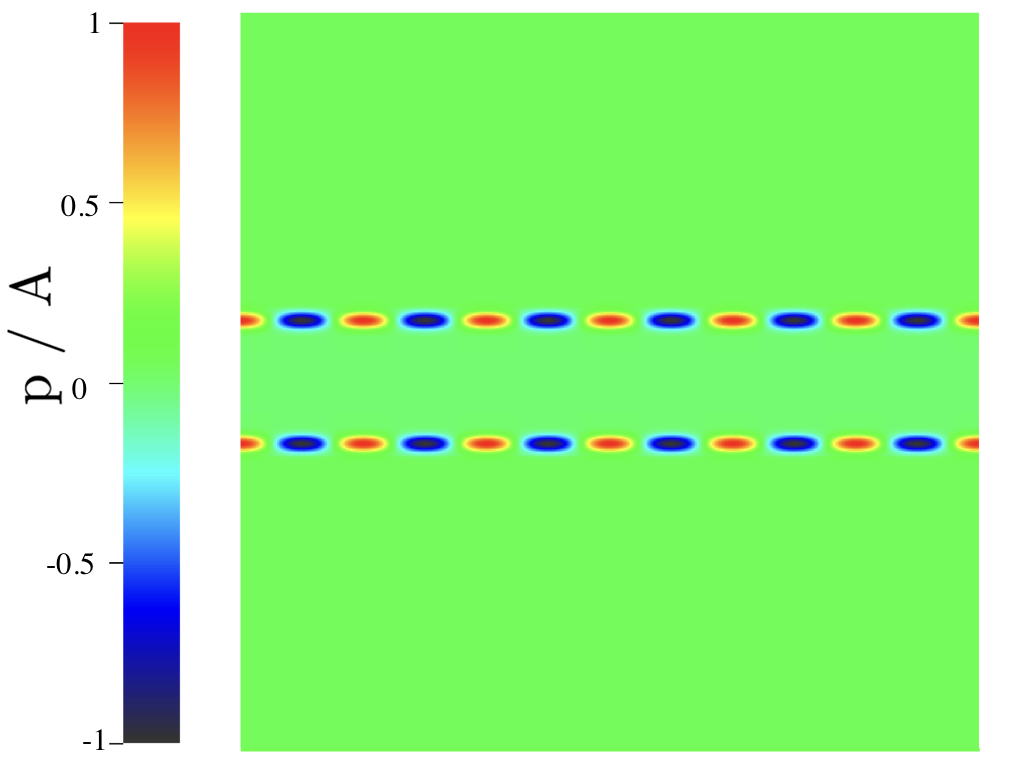}
 \caption{The initial setup of our simulations of a flowing filament inside a hot halo. (top): density contrast of the simulation, $\delta_{\rm{f}}=\rho / \rho_{\rm{fil}}$ (bottom): pressure distribution in units of the pressure perturbation amplitude A. The filament and halo background are in pressure equilibrium and initial pressure perturbations with wavelength $\lambda$ are placed along the interface between the two media.
 }
 \label{fig:initCond}
\end{figure}

In this Section, we describe the numerical simulations used to investigate the effect of nonlinear KHI on cold accreting filaments using idealised simulations. Similar to \cite{Mandelker2016} we simulate a gas slab with high density and low temperature, representing a cold accreting filament flowing through a hot, less dense background medium, representing the halo gas, see Fig. \ref{fig:initCond}. To investigate the nonlinear evolution of KHI, we place pressure perturbations on the interface between the two media.

Our simulations are run in 2D with the Eulerian code RAMSES \citep{ramses}. The piecewise linear reconstruction used in the Godunov scheme is the MonCen slope limiter \citep{Leer1977}, and a HLLC approximate is used for the Riemann solver \citep{Toro1994}. The domain of the simulation lays in the xy plane and is a square box. The filament is centred at the middle of the box and has a radius of $R_{s}=1/12L$ where $L$ is the box size. We use periodic boundary conditions at $y = 0$ and $y=L$ and outflow boundary conditions at $x=0$ and $x=L$. 
Outflow boundaries are chosen in order to minimise waves reflected off the box sides. Gas that crosses the boundary is lost. We have checked that and found the amount of gas lost during simulations is negligible.

The halo gas, at $|x|>R_s$ is initially set to have a density $\rho_{\rm{halo}} = 1.3 \times 10^{-4} \;\rm{cm^{-3}}$
and it is static ($v_{\rm{halo,0}}=0$). 
However, because our simulations do not include cooling they are scale-free with respect to length, time, temperature and density, see eq. \ref{eq:transformation} for further details.

The filament gas at $|x|<R_s$ is set to a density $\rho_{\rm{fil}} = \delta  \rho_{\rm{halo}}$. $\delta=152$ is the density contrast between the halo and the filament and is kept fixed for all our simulations. 
As commonly done in the literature, we use Mach number instead of a physical velocity for the filament, also because KHI is expected to be scale free. The Mach number $M_b$ is unitless, and it is defined as the ratio between the filament velocity and the sound speed in the halo background:

\begin{equation}
M_b = \frac{v_{\rm{fil}}}{c_{\rm{halo}}}
\end{equation}

where $v_{\rm{fil}}$ is the velocity of the filament and $c_{\rm{halo}}$ is the sound speed of the halo.
In setting our initial conditions we closely follow the approach of \cite{Mandelker2016}.

In order to reduce numerical noise and to approximate real filaments where the boundary is not perfectly discontinuous, we smooth the velocity and density across the interface between the halo and the filament using a ramp function: 
 
\begin{equation}
\begin{split}
f(x) &=f_b + 0.25(f_s - f_b) \times \\
& \bigg [ 1 + \rm{tanh}\bigg (\frac{R_s-x}{\sigma}\bigg ) \bigg ] \bigg [ 1 + \rm{tanh}\bigg (\frac{R_s+x}{\sigma}\bigg ) \bigg ] .\
\end{split}
\end{equation}

This results in a smooth interface between the two fluids. $\sigma$ represents what we will call the interface thickness. Perturbations of wavelengths of similar size or smaller than $3 \sigma$ are suppressed by this smooth interface \citep{Robertson2010}. 
We have assumed both the filament and the halo background to be ideal gases with adiabatic index $\gamma = 5/3$.
We initially assume pressure equilibrium between the filament and halo gas. The value of the pressure is given by the temperature and density through the ideal gas law. The temperature of the halo (filament) is $T_{\rm{halo}}=2.3 \times 10^6 \;\rm{K}$ ($T_{\rm{fil}}=1.5 \times10^4 \; \rm{K}$). Pressure perturbations are placed along the filament interface on both sides: 

\begin{equation}
P = A \rm{cos}\bigg (\frac{2 \pi}{\lambda} y \bigg ) \times \bigg [ \rm{exp}\bigg(- \frac{(x-R_s)^2}{2 \Sigma^2} \bigg) + \rm{exp}\bigg(-\frac{(x+R_s)^2}{2 \Sigma^2}\bigg)\bigg],\
\end{equation}

where $A = 0.05$ is the pressure amplitude, $\Sigma = 2\sigma$ is the width of the perturbation and $\lambda$ the perturbation wavelength. We have run further simulations for different values of $\Sigma$ and found no significant differences. Others have also found the exact value of $\Sigma$ has negligible impact as long as it is larger than $\sigma$.

We use a single grid with a fixed resolution. 
In order to keep possible resolution effects independent of varying perturbation wavelengths, we fix the number of cells per wavelength $\lambda$. The cell size, $\Delta$, is $2^{-13}$, $2^{-12}$, $2^{-11}$ for $\lambda / R_s = 1, 2, 4$ respectively.

In all our fiducial calculations we set the filament radius, $R_s=4$ kpc and $L=48$ kpc. 
We have run all our simulations for 1.6 Gyr assuming $T_{\rm{fil}}=1.5\times10^{4} \;\rm{K}$. Because the adiabatic simulations (and the development of Kelvin-Helmholtz instability) that we are performing are scale free \citep{Chandrasekhar1961}, this physical time may be rescaled using the relations \ref{eq:transformation} above.
As a comparison, the Kelvin-Helmholtz time $t_{KH}$:

\begin{figure*}
\minipage{0.5\textwidth}
  \includegraphics[width=\linewidth]{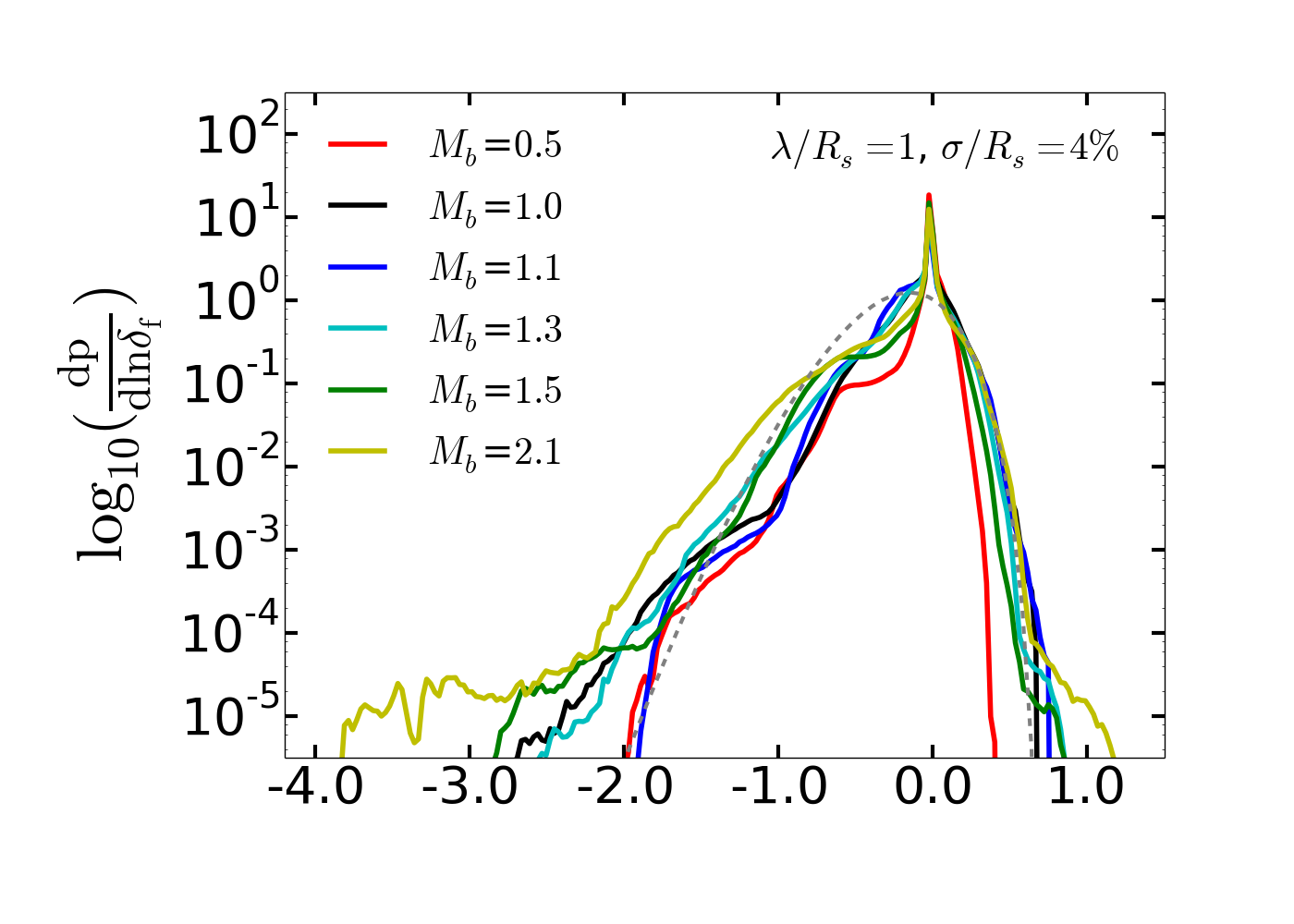}
\endminipage
\minipage{0.5\textwidth}
  \includegraphics[width=\linewidth]{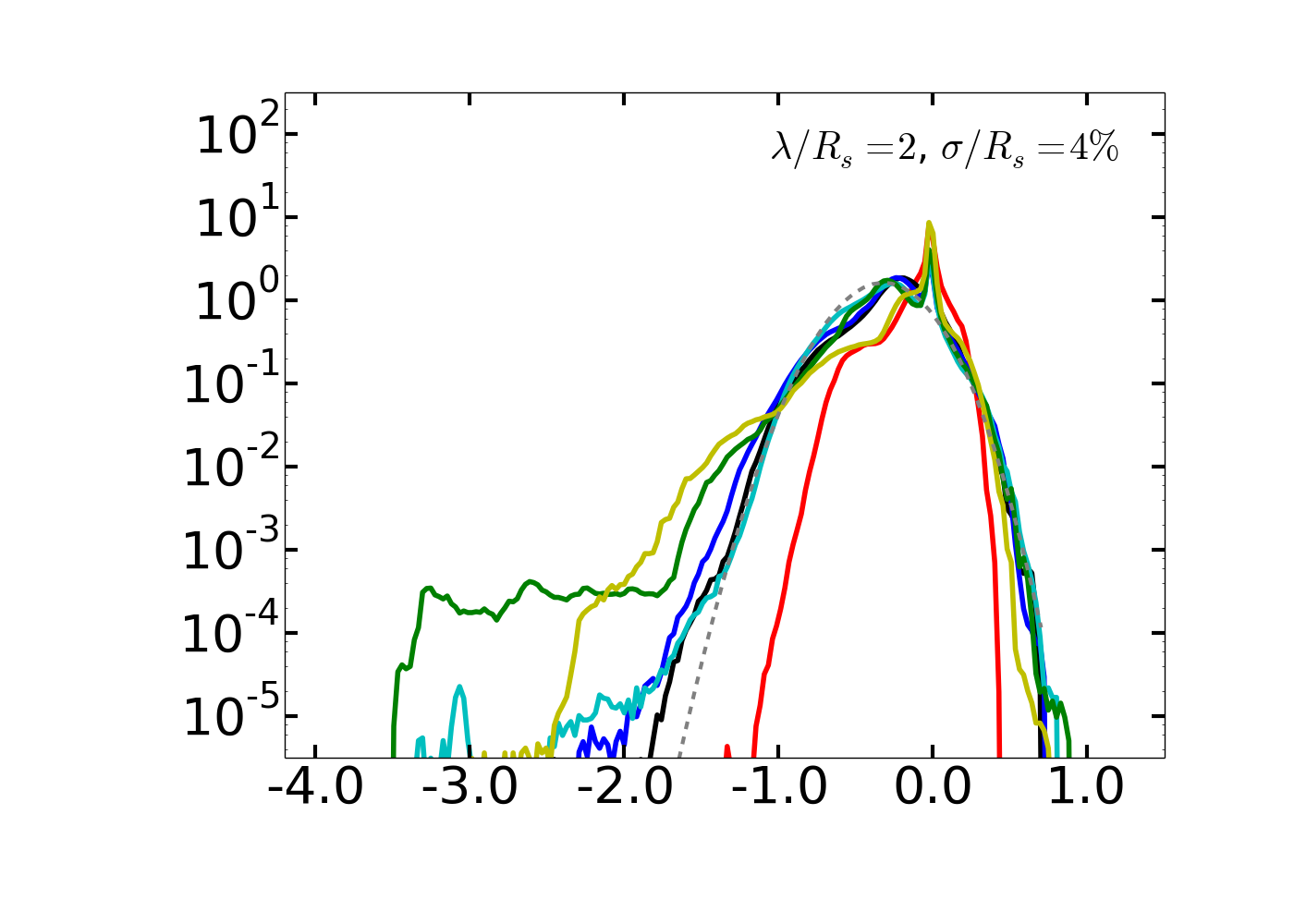}
\endminipage\hfill
\minipage{0.5\textwidth}%
  \includegraphics[width=\linewidth]{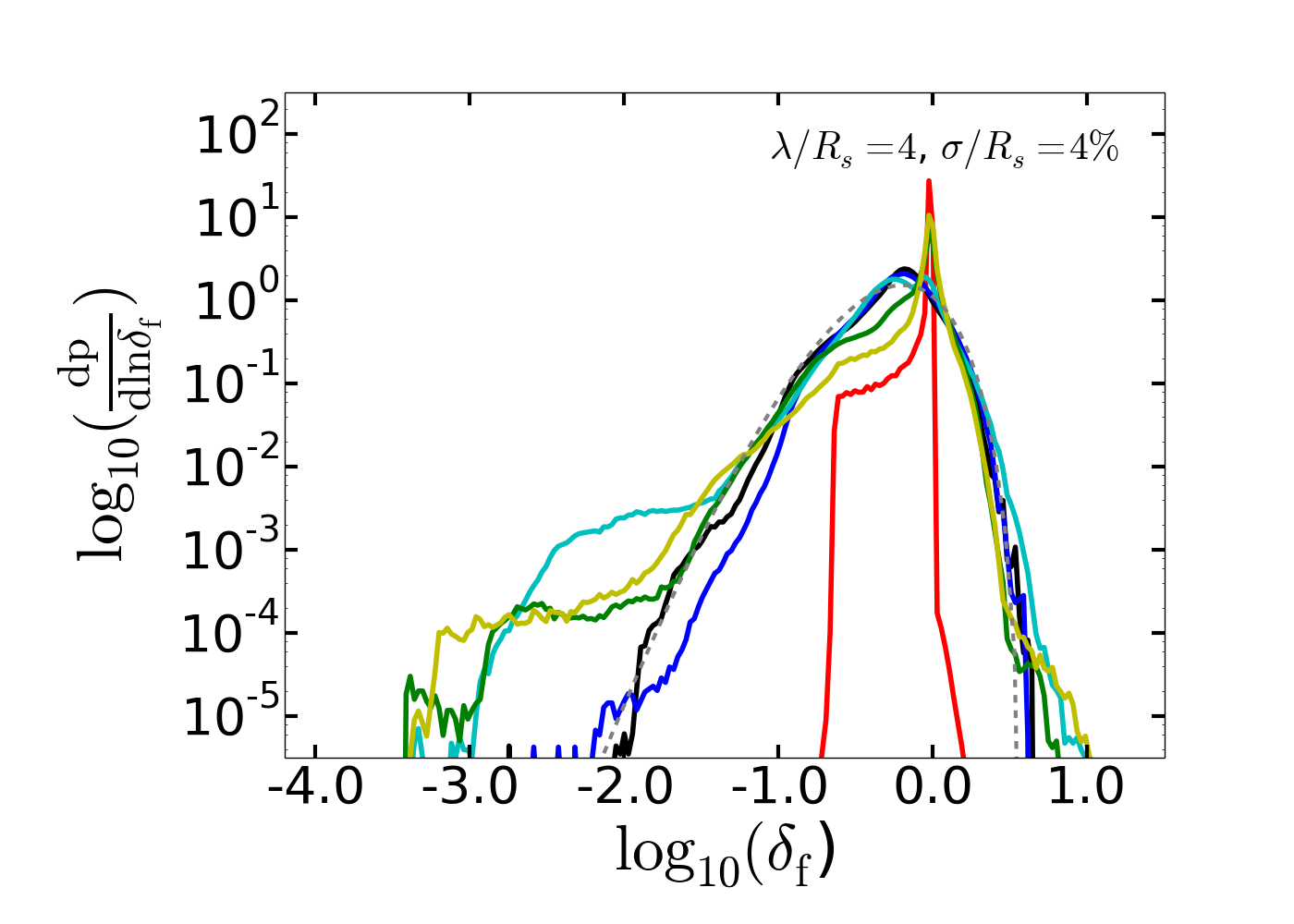}
\endminipage
\minipage{0.5\textwidth}%
  \includegraphics[width=\linewidth]{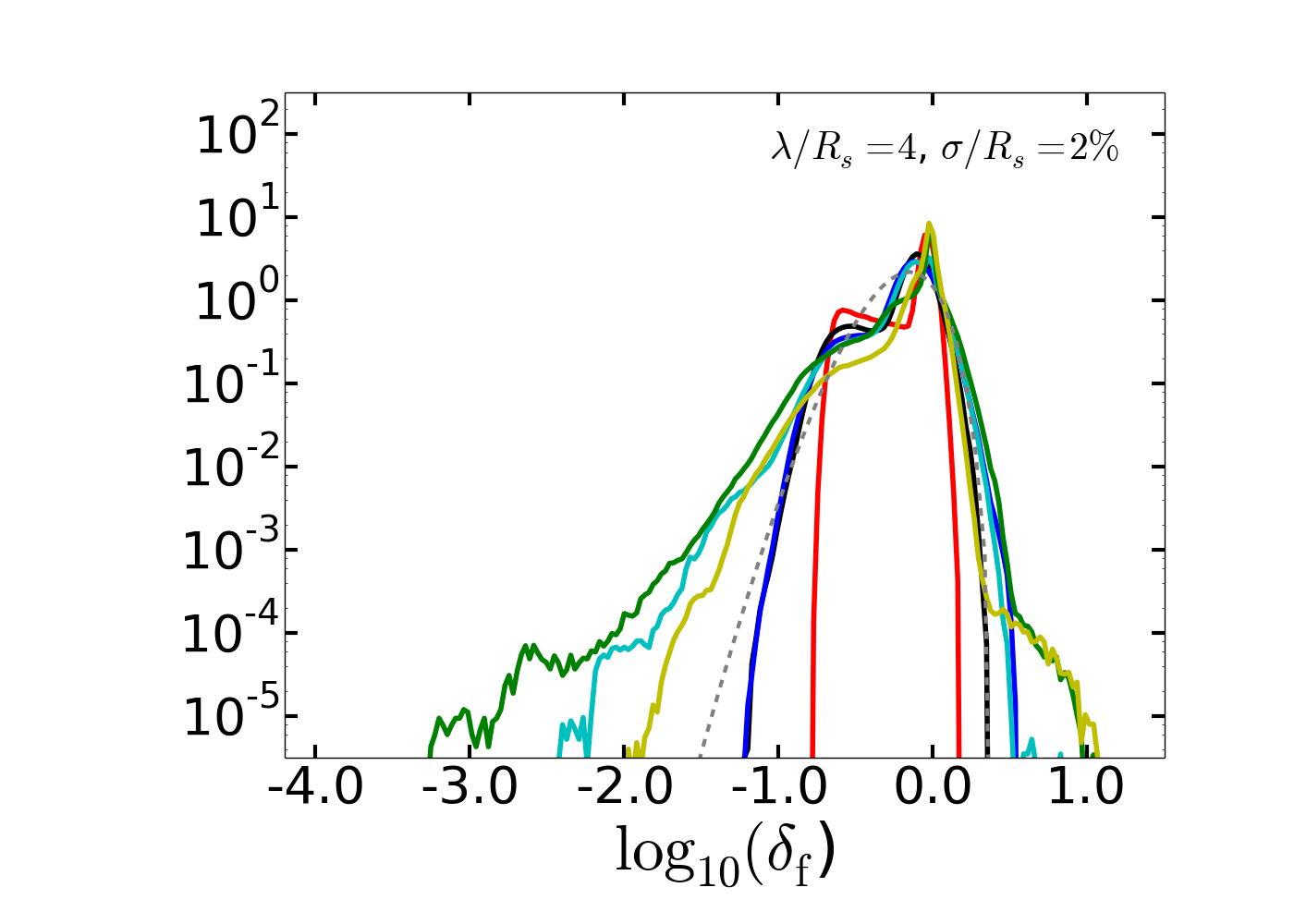}
\endminipage
\caption{Time averaged gas density distribution (PDF) of cold gas ($< 5 \times 10^4$ K) for our four fiducial runs (see Table \ref{table_initCond}). Each panel shows the results for $M_b=0.5$ to $M_b=2.1$. All other parameters (wavelength: $\lambda/R_s$ and interface thickness between the filament and halo: $\sigma/R_s$) are kept constant for each set of simulations. The average is taken over very long timescales with respect to the expected timescale for KHI to develop. See text for further details. The dashed lines correspond to a skewed log-normal distribution overlaid to fit simulations with $M_b=1.0$. The parameters for (S,T) for the skewed log-normal fits are as follows: 0.07, 0.04 (top left), 0.06, 0.005 (top right), 0.073, 0.045 (bottom left), 0.035, 0.03 (bottom right).}
\label{fig:MbPDFs}
\end{figure*}

\begin{equation}
t_{KH}=\frac{\sqrt{\delta}}{M_b}\frac{\lambda}{c_b} 
\sim 300 \Bigg ( \frac{\lambda}{R_s} \Bigg ) \Bigg ( \frac{R_s}{4 \textrm{kpc}} \Bigg ) \Bigg ( \frac{\sqrt{\delta} }{M_b} \Bigg ) \Bigg (\frac{T}{10^6 \textrm{K}} \Bigg )^{-1/2} \;\textrm{Myr}
\end{equation}

for the same set of parameters and $M_b=1.0$, $\lambda=2 R_s$ is 165 Myr.

We have run several simulations with different initial conditions shown in table \ref{table_initCond} and have chosen all our initial conditions to meet the requirement $\lambda > 30 \sigma > 60 \Delta$ stated by \cite{Mandelker2016}, for which they find good convergence. The range of $M_b$ was chosen to fit likely velocities of the infalling filament through a hot halo. $M_b=0.5$ ($M_b=2.1$) corresponds to a filament flowing at $1/4$ (1) times the freefall velocity assuming the halo is at the virial temperature.

\begin{table}

\centering
  \caption{Parameters used for initial conditions in our simulations. $\lambda/R_s$ is the wavelength of the perturbation at the interface between the filament and hot halo. $\sigma / R_s$ is the thickness of the interface between the two media. $M_b$ is the Mach number with respect to the sound speed of the halo gas. In particular, the values for the range 0.5 - 2.1 are $M_b=$ 0.5, 1.0, 1.1, 1.3, 1.5, 2.1. Simulations with parameters above the horizontal line belong to our fiducial runs.}
  \label{table_initCond}
\begin{tabular}{| l | c | r | }
$\lambda / R_s$ & $\sigma / R_s$   & $M_b$  \\
 \hline
  4 & 2\% & 0.5 - 2.1 \\
 1 & 4\% & 0.5 - 2.1 \\
 2 & 4\% & 0.5 - 2.1 \\
 4 & 4\% & 0.5 - 2.1 \\
 \hline
 1 & 9\% & 1.1, 2.1\\
 2 & 9\% & 1.1, 2.1 \\
 4 & 9\% & 1.1, 2.1 \\
 2 & 2\% & 1.1 \\
 
\end{tabular}
\end{table}

 \begin{figure*}
 \minipage{0.5\textwidth}
  \includegraphics[width=\linewidth]{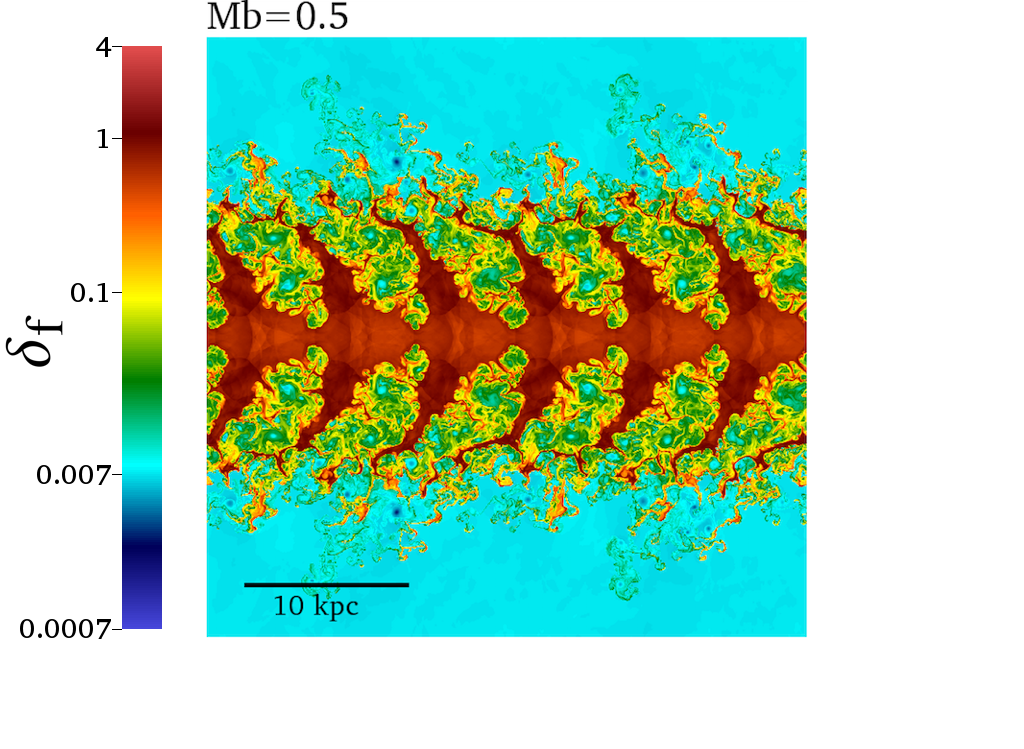}
\endminipage
\minipage{0.5\textwidth}%
  \includegraphics[width=\linewidth]{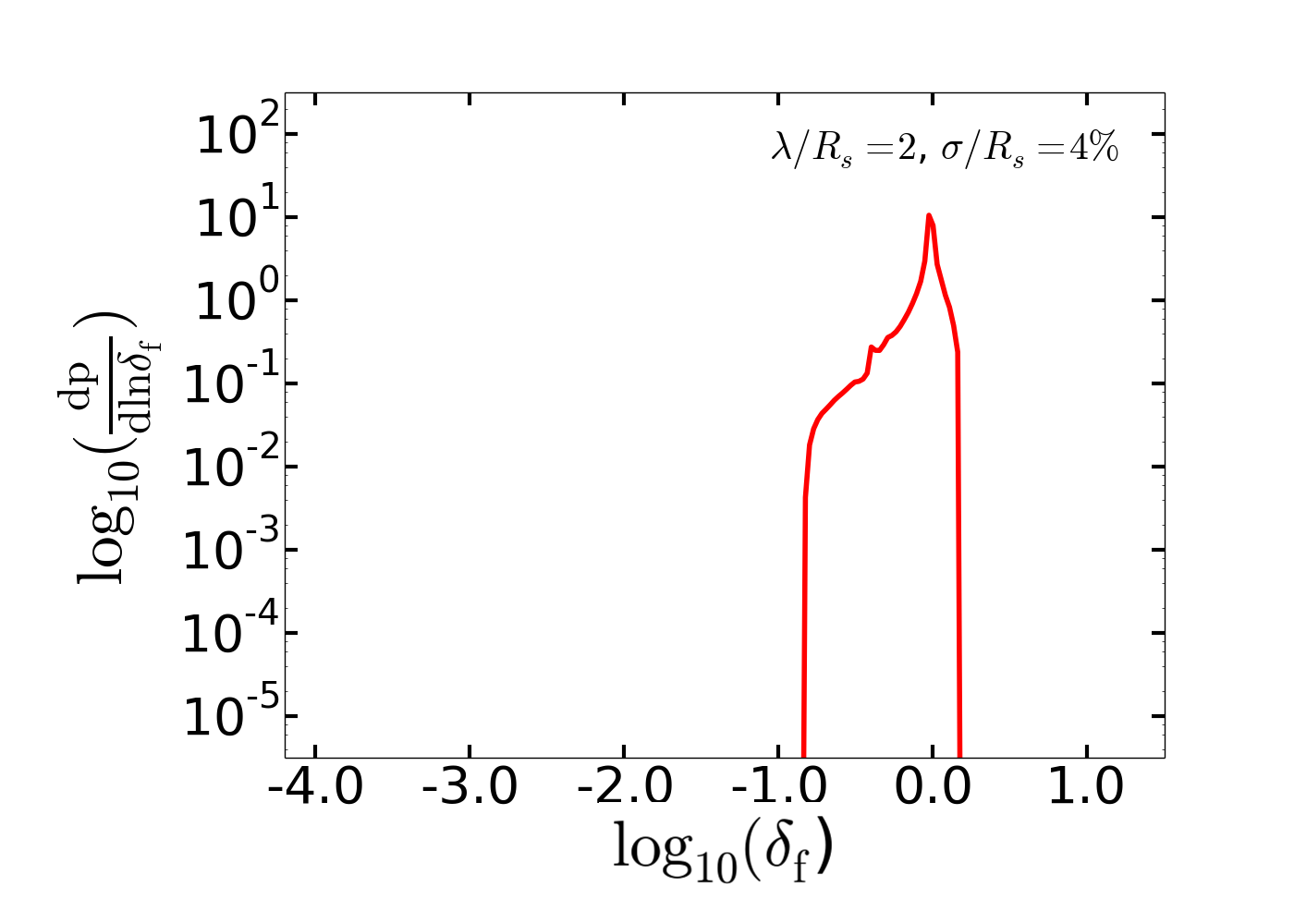}
  \endminipage\hfill
\minipage{0.5\textwidth}
  \includegraphics[width=\linewidth]{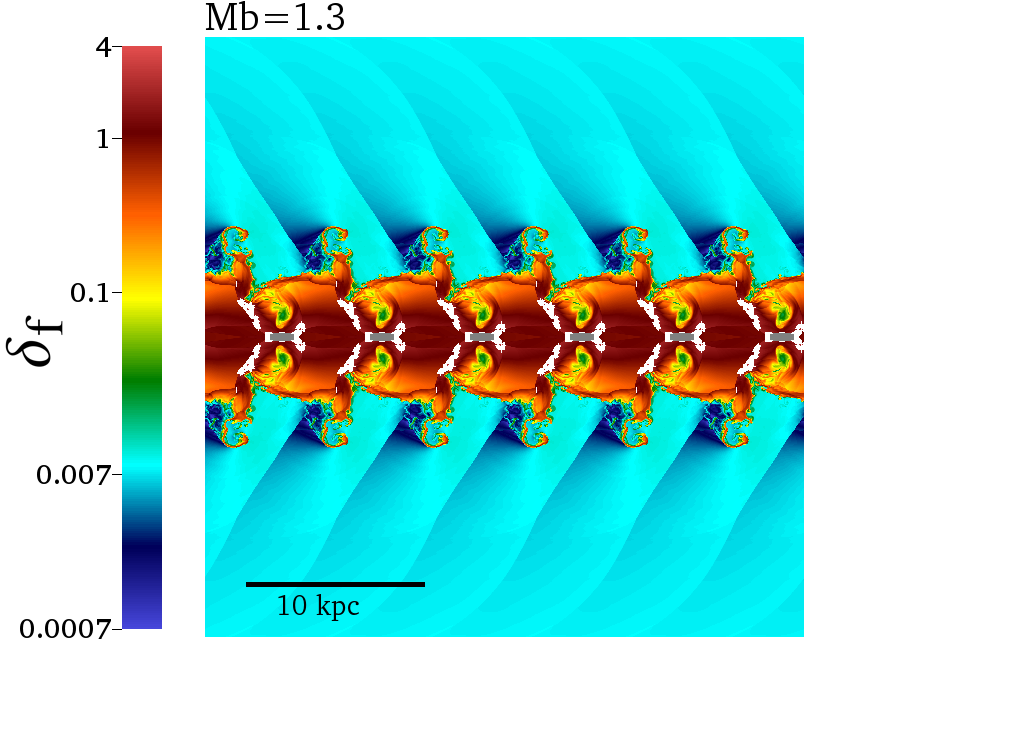}
\endminipage
\minipage{0.5\textwidth}%
  \includegraphics[width=\linewidth]{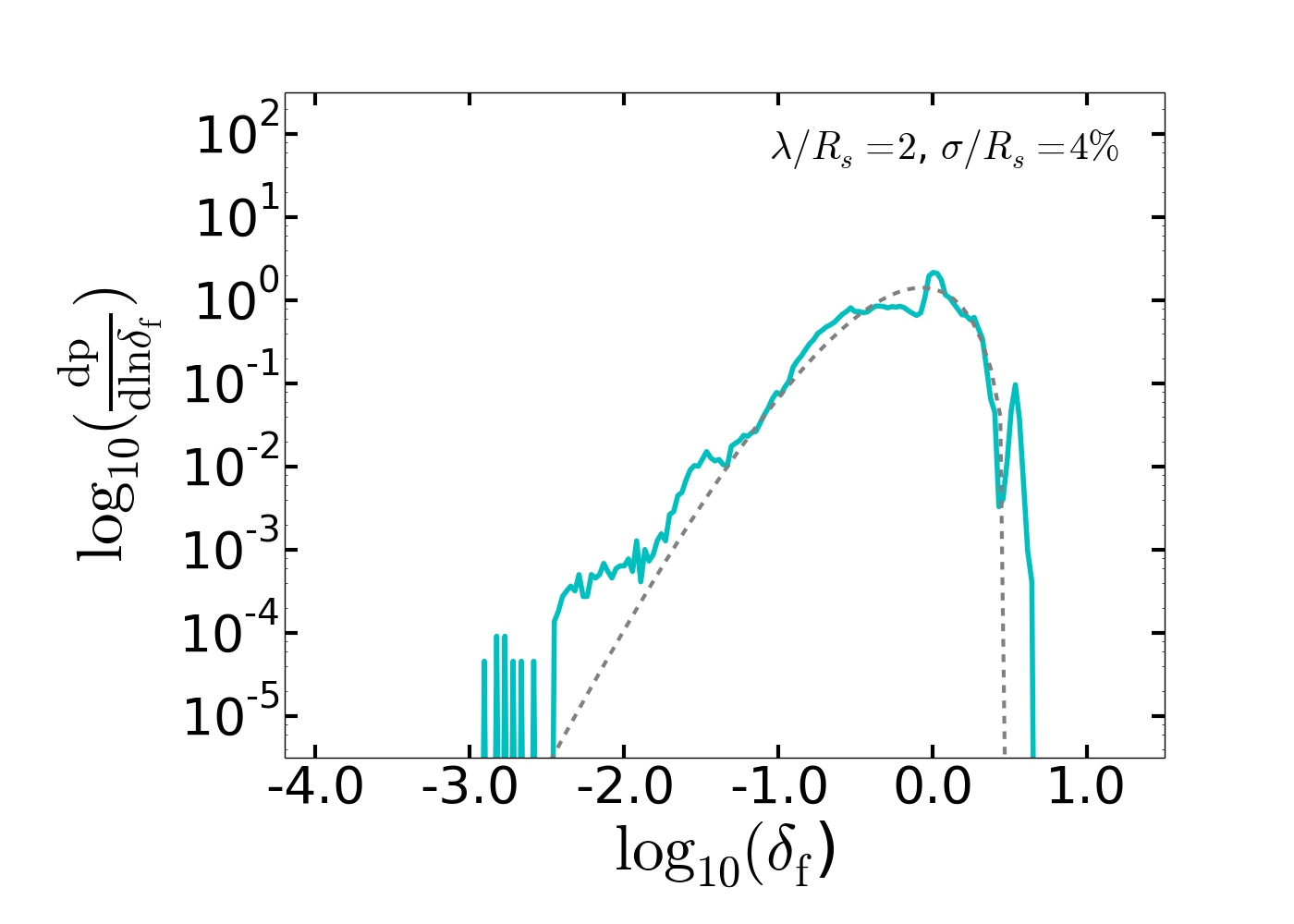}
\endminipage\hfill
\minipage{0.5\textwidth}%
  \includegraphics[width=\linewidth]{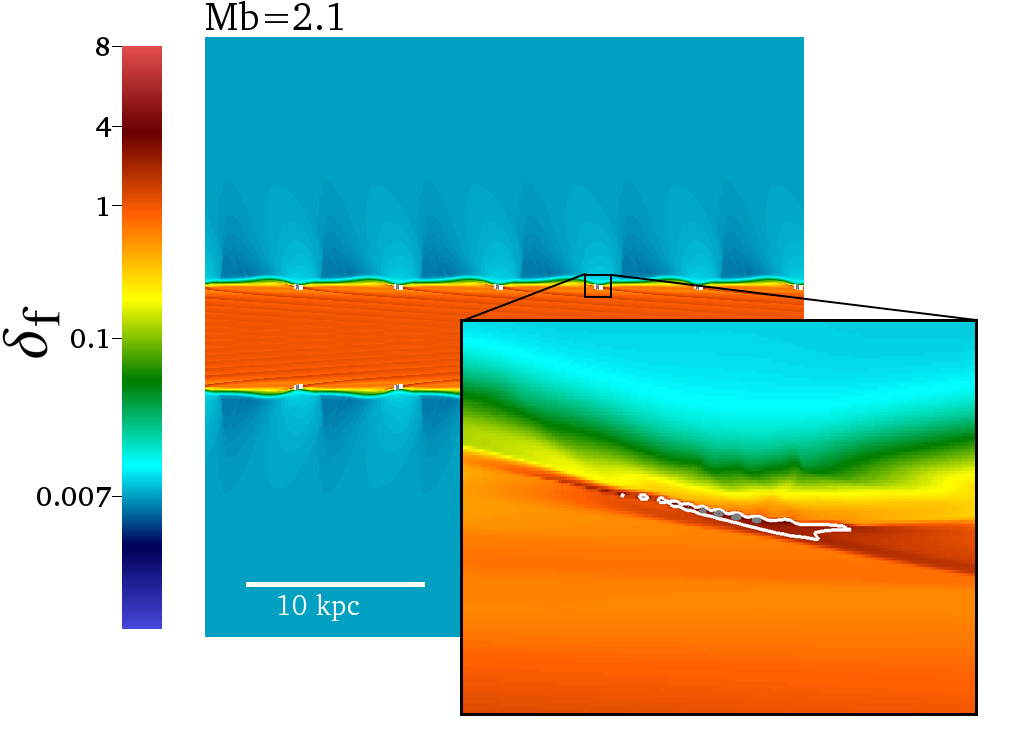}
\endminipage
\minipage{0.5\textwidth}%
  \includegraphics[width=\linewidth]{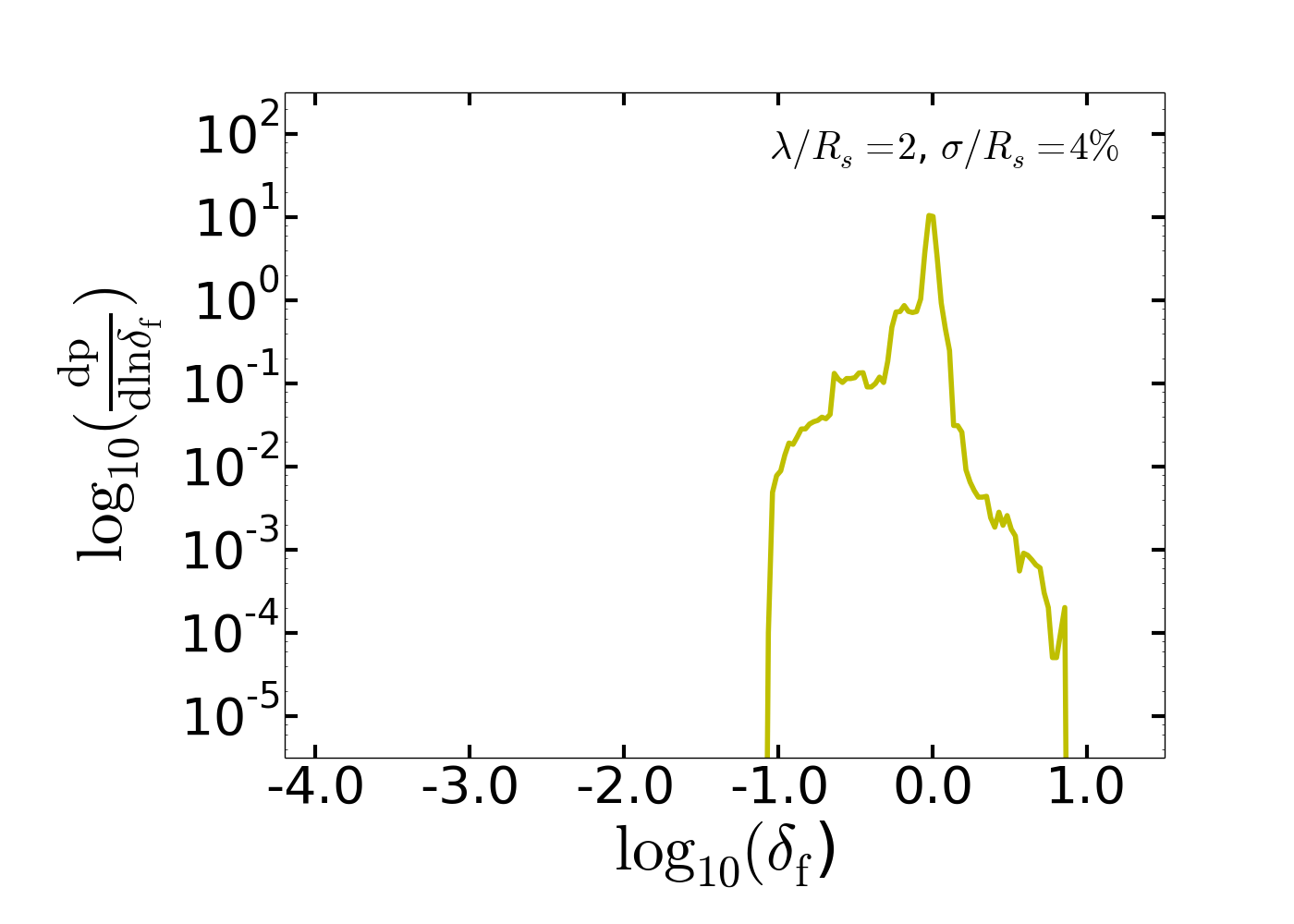}
\endminipage
 \caption{Snapshots (left) \& PDF (right) of some of our simulations at relevant times during the evolution of the filament. Results are shown for $M_b = 0.5$ (top), $M_b=1.3$ (middle), $M_b=2.1$ (bottom) after a time of $\sim 4 t_{KH}$ (top and middle) and $\sim 8 t_{KH}$ (bottom). 
 All simulations have the exact same initial conditions ($\lambda/R_s = 2$ and $\sigma/R_s=4\%$) and are identical apart from their initial speed of the filament. The parameter displayed is the density contrast with respect to the initial density of the filament. The colorbar ranges from $ \delta_{\rm{f}} = 0.0007 = 0.1 \rho_{\rm{fil}}$ to $\sim 4 \delta_{\rm{f}} = 4  \rho_{\rm{fil}}$ ($\sim 8 \delta_{\rm{f}} = 8  \rho_{\rm{fil}}$ for the last image). Density contours show regions of high densities. White: $2 \rho_{\rm{fil}}$ Grey: $3\rho_{\rm{fil}}$. The zoom in the bottom left shows a high density filamentary structure, resembling "chains" of very dense cold gas developing at the filament-halo interface.}
 \label{fig:Mb1.3-extreme}
 \end{figure*}

Because of the scale-free nature of our methods, the results of the simulations will be equivalent 
for a set of different initial conditions related by the following transformations (at a fixed $M_b$, $\delta_{\rm{f}}$ and $\lambda/R_s$):

\begin{equation} 
\label{eq:transformation}
\begin{aligned}
& T_{\rm{halo}}' = \alpha T_{\rm{halo}} \\
& T_{\rm{fil}}' = \alpha T_{\rm{fil}}  \\
& v_{\rm{fil}}' = \alpha^{1/2} v_{\rm{fil}} \\
& c_{\rm{halo}}' = \alpha^{1/2} c_{\rm{halo}} \\
& R_s' = \beta R_s \\
& t' =t \beta \alpha^{-1/2}\\
& \rho_{\rm{halo}}'=\gamma \rho_{\rm{halo}}\\
& \rho_{\rm{fil}}'=\gamma \rho_{\rm{fil}}.
 \end{aligned}
\end{equation}

In the above transformations, $R_s$ represents the radius of the filament, $T_{\rm{halo}}$ is the halo temperature, while $\alpha$, $\beta$ and $\gamma$ are rescaling parameters.

\section{Results}
\label{sec_results}

\subsection{Development of high densities during filament accretion}
\label{sec:devHighDensity}
In this Section, we investigate whether accreting filaments into hot haloes develop high density regions during their lifetime (the time they need to reach the central galaxy,  $t_{\rm{cross}}$) and if so, which properties of the filament will most likely lead to these structures. In particular, we will address the following questions: 
i) what are the density distributions of cold accreting filaments, 
ii) what are the largest densities that develop,
iii) how do these high density regions evolve, and
iiii) what are the morphological properties of the densest clumps. 
We will explore these questions by varying the properties of the filament in terms of velocity, initial perturbation wavelength and interface thickness 
(see Section \ref{sec_methods}).

\subsubsection*{Density distribution of accreting filaments}

In Fig. \ref{fig:MbPDFs}, we present the time averaged probability distribution functions (PDFs) of the cold gas ($T<5\times10^4$ K)
\footnote{
Note, that this breaks the scale invariance, though only with respect to parameter $\alpha$ in eq. \ref{eq:transformation}} densities within the computational volume. 
The time average is performed from 2 $t_{\rm{KH}}$ until $1.6$ Gyr,
that represents a period long enough to include the expected nonlinear development of KHI and is of similar order as the halo crossing time, $t_{\rm{cross}}$ (see Section \ref{sec_methods}). 
The densities are expressed as a density contrast ($\delta_{\rm{f}}$) with respect to the initial filament density ($\rho_{\rm{f,init}}$). 
There are four sets of simulations, each with its own fixed $\lambda / R_s$ and $\sigma / R_s$ and the same variety of Mach numbers (see Table \ref{table_initCond}), represented by the different solid lines in the figure. This combination of simulations allows us to understand which parameter drives the development of high densities. 

\begin{figure*}
\minipage{0.5\textwidth}
  \includegraphics[width=\linewidth]{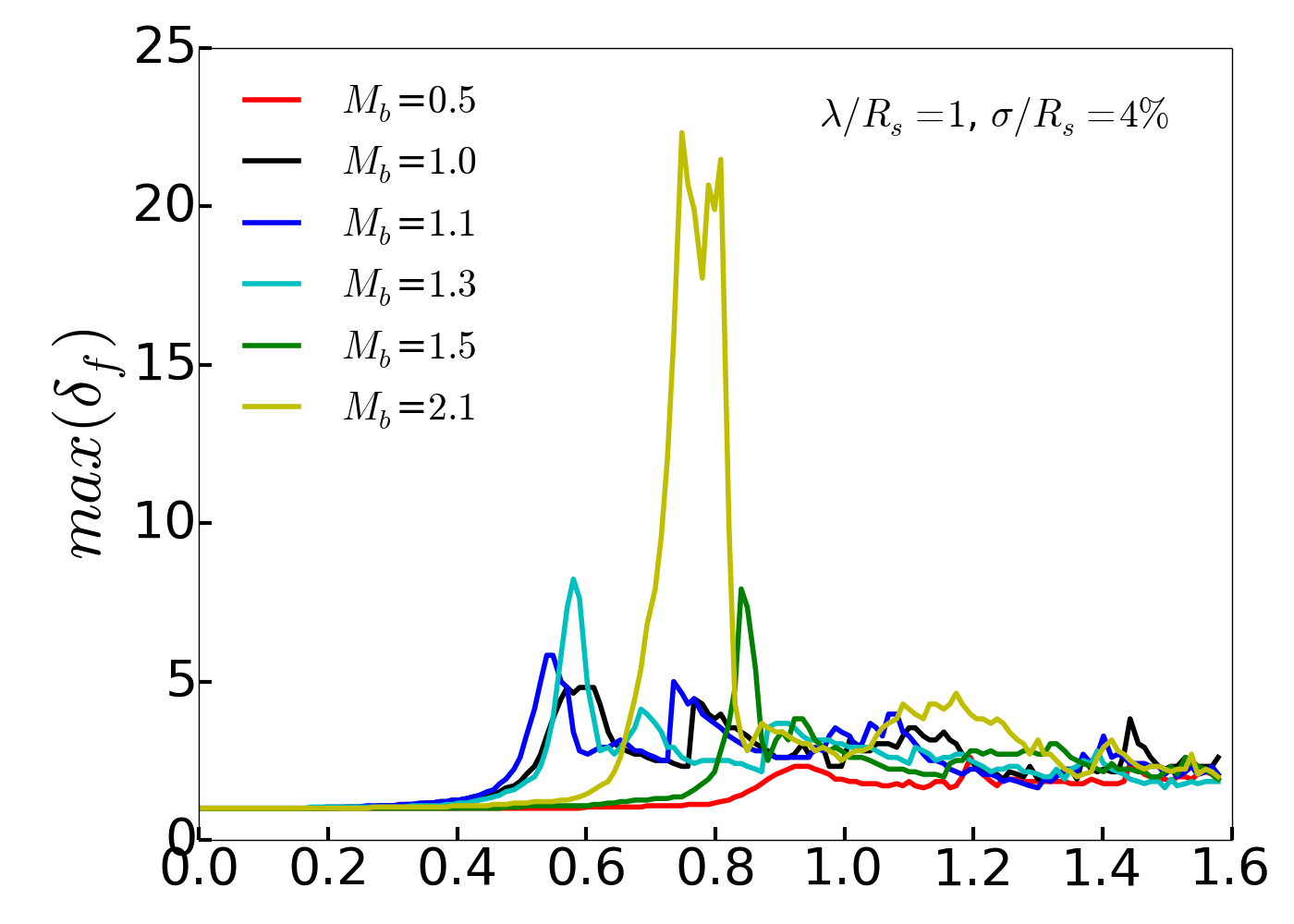}
\endminipage
\minipage{0.5\textwidth}
  \includegraphics[width=\linewidth]{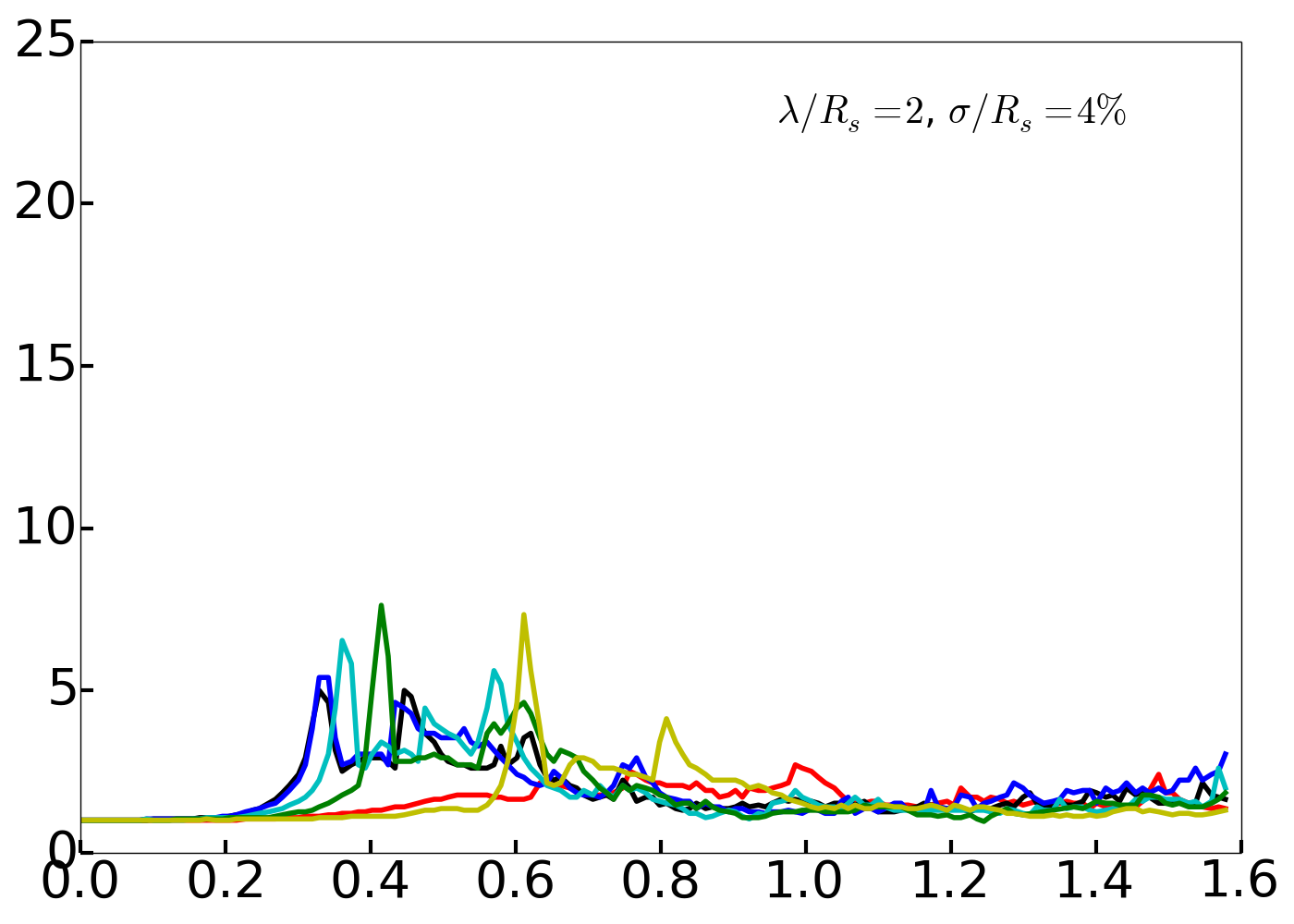}
\endminipage\hfill
\minipage{0.5\textwidth}%
  \includegraphics[width=\linewidth]{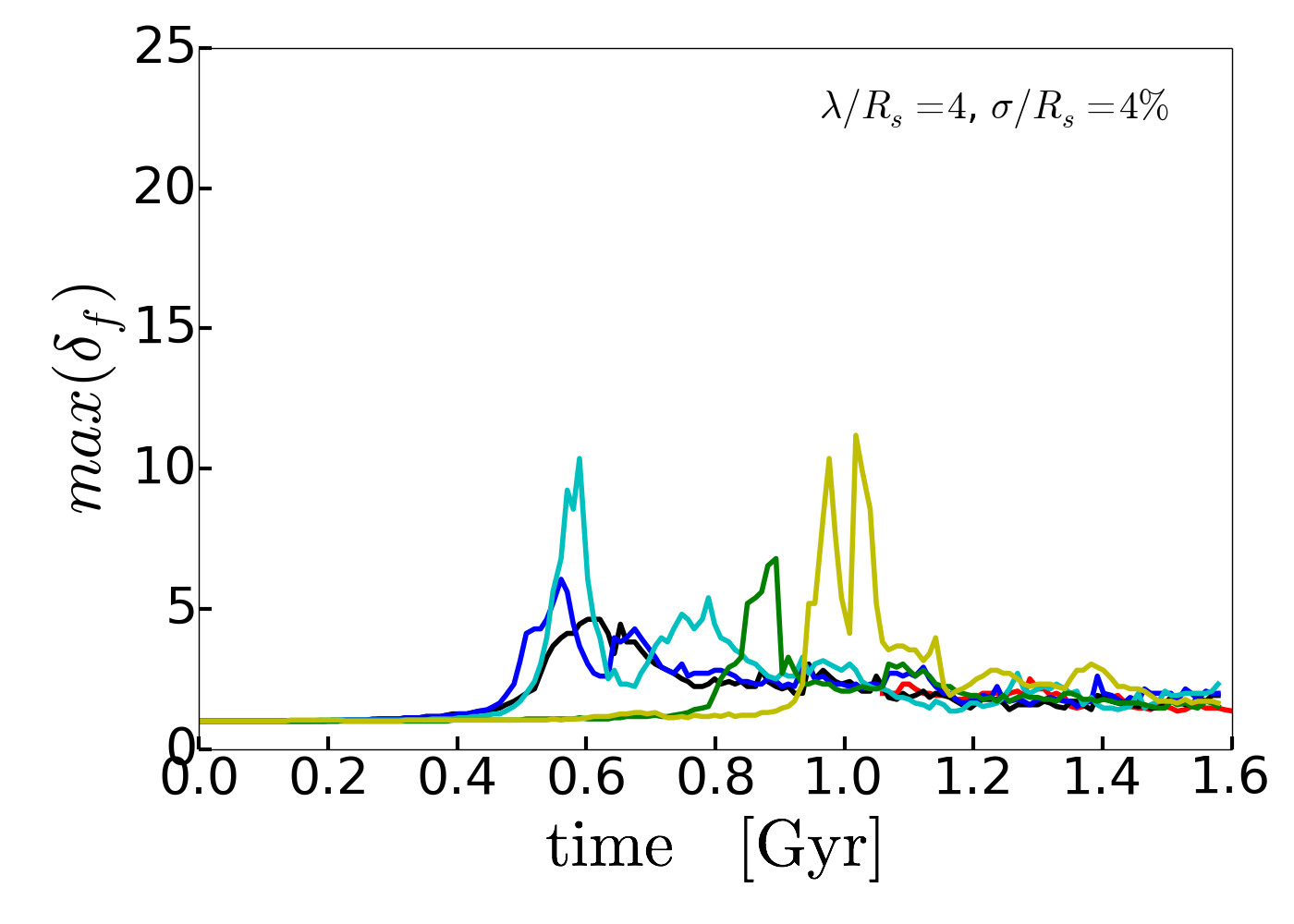}
\endminipage
\minipage{0.5\textwidth}%
  \includegraphics[width=\linewidth]{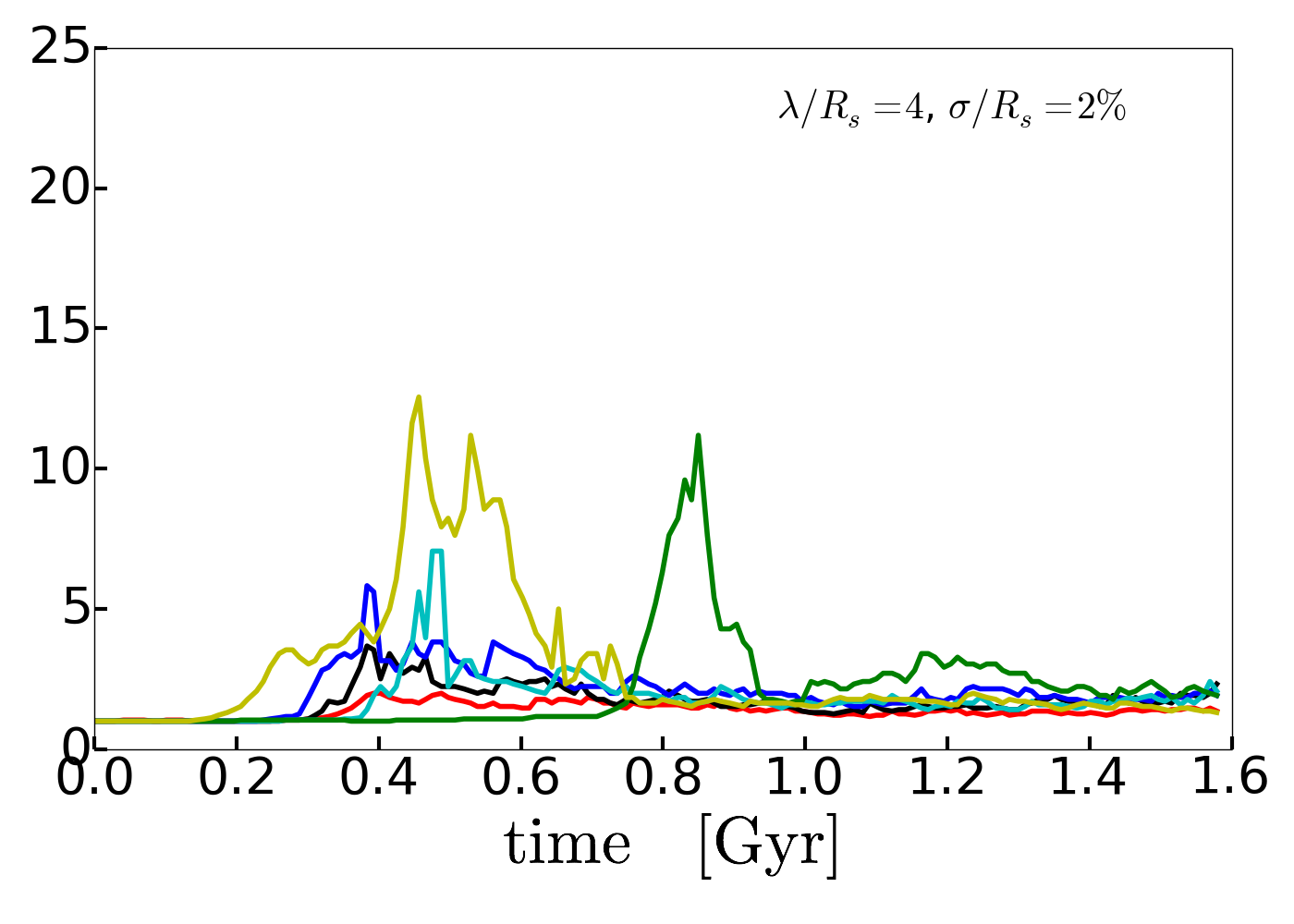}
\endminipage

 \caption{Maximum density contrast as a function of time for a range of $M_b$ for our fiducial runs. The time axis is normalised to $R_s=4$ kpc (see equation \ref{eq:transformation} for rescaling).}
 \label{fig:2RsvaryMb184__maxDensityWithTime}
 \end{figure*}
We find that the time averaged PDFs of filaments with $M_b\approx1$
are qualitatively consistent with skewed log-normal distributions that are overlaid for illustrative purposes with dashed lines in Figure \ref{fig:MbPDFs}. 
These skewed log-normals are defined as in \cite{Hopkins2013b}:

\begin{equation} \label{eq:lnPDF}
\begin{aligned}
\textrm{PDF} (\textrm{ln} \rho)d \textrm{ln}  \rho &= I_1(2 \sqrt{\lambda u}) \textrm{exp}[-(\lambda+u)] \sqrt{\frac{\lambda}{u}} \textrm{d}u \\
u &\equiv \frac{\lambda}{1+T} - \frac{\textrm{ln} \rho}{T} \quad\quad (u\geq 0) \\
\lambda &\equiv \frac{S}{2T^2}
\end{aligned}
\end{equation}

where $I_1(x)$ is the modified Bessel function of the first kind and PDF $= 0$ for $u < 0$. When $T \rightarrow 0$ the above distribution converges to a log-normal distribution:

\begin{equation}
\label{eq:PDF_lognormal}
\textrm{PDF} (\rho)d \rho= \frac{1}{\rho \sigma_s \sqrt{2 \pi}} 
\textrm{exp} \Bigg ( - \frac{(\textrm{ln}( \rho ) - \mu )^{2}}{2 \sigma_{s}^2}\Bigg ) 
d\rho,
\end{equation}

where $\mu$ is the average of the logarithm of the density.
Simulations with other $M_b$ are also in qualitative agreement with skewed log-normals but show additional components. In Section \ref{sec_discussion}, we will discuss the possible origin
of these distributions. 
We find three additional components deviating from the skewed log-normal: a low density tail, a high density tail approximating a power law and a peak at $\delta_{\rm{f}}=1$. 
Most simulations with Mach Numbers $M_b > 1.3$ show the broadest cold gas density distribution, deviating from the skewed log-normal with both a high and low density tail. In particular, these simulations have cold gas reaching very high (low) values of more (less) than 20 ($10^{-3}$) the initial filament density. 
The high density tail is most pronounced for $\lambda/R_s=4$, $\sigma/R_s=2 \%$ and $M_b=$1.5, 2.1 (lower right figure) and seems to follow a power law. It is present in each simulation with $M_b>1.3$, however not always well visible in Fig. \ref{fig:MbPDFs}. This is because PDFs presented in Fig. \ref{fig:MbPDFs} are time averaged and therefore short-lived high-density regions will not contribute substantially to the PDF. Furthermore, Fig. \ref{fig:Mb1.3-extreme} (bottom right) displays this seemingly missing high density tail for the case of $\lambda/R_s = 2$, $\sigma/R_s=4 \%$ at an individual time. 
Further, all sets of simulations show a peak at $\delta_{\rm{f}} = 1$, which corresponds to unperturbed cold gas of the filament. It is present for all simulations with $M_b = 0.5$, and it disappears for runs with $M_b \approx 1.0$, $\lambda/R_s > 1$ and then reappears when $M_b > 1.3$.
Lastly, we note that the subsonic case ($M_b=0.5$) results in the narrowest PDF that never exceeds $ \delta_{\rm{f}}=3$ in any of our simulations.

On the left of Fig. \ref{fig:Mb1.3-extreme}, we show for illustrative purposes three simulation snapshots showing the density contrast ($\delta_{\rm{f}}$) for $M_b=0.5$ (top), $M_b=1.3$ (middle) and $M_b=2.1$ (bottom)
at $t=4\times t_{\rm{KH}}$ for the two upper images and at $t=8\times t_{\rm{KH}}$ for $M_b=2.1$. The regions with density contrasts $\delta_{\rm{f}} \geq 3$ ($\delta_{\rm{f}} \geq 2$) are indicated by grey (white) contours. These simulations present very different PDFs at the time of the snapshot (Fig. \ref{fig:Mb1.3-extreme} right).
The slower moving filament ($M_b=0.5$) shows a flatter density as a function of vertical distance and does not develop high densities. The faster moving filament ($M_b=1.3$) develops higher densities at the center and has a broader PDF. The fastest filament with $M_b=2.1$ develops the highest densities, which are produced at the interface between the two media.

In addition to the time averaged density distribution we are also interested in the short-lived high density regions because these regions might become long lived high density regions, due to additional physical processes, which will be discussed in Section \ref{sec_discussion}. Indeed, we find these regions exist for some filaments, as shown in \ref{fig:Mb1.3-extreme}. 
In order to capture the development of high-density regions at a given time and the general time evolution of the PDF, 
we decided to use as a possible proxy for these properties: the maximum density contrast $\rm{max}(\delta_{\rm{f}})$. 
This parameter may be sensitive to numerical noise but, as we show later on, this is not the case in our simulations (see Fig. \ref{fig:seedSnapshot}).

\begin{figure} 
 \includegraphics[width=\columnwidth]{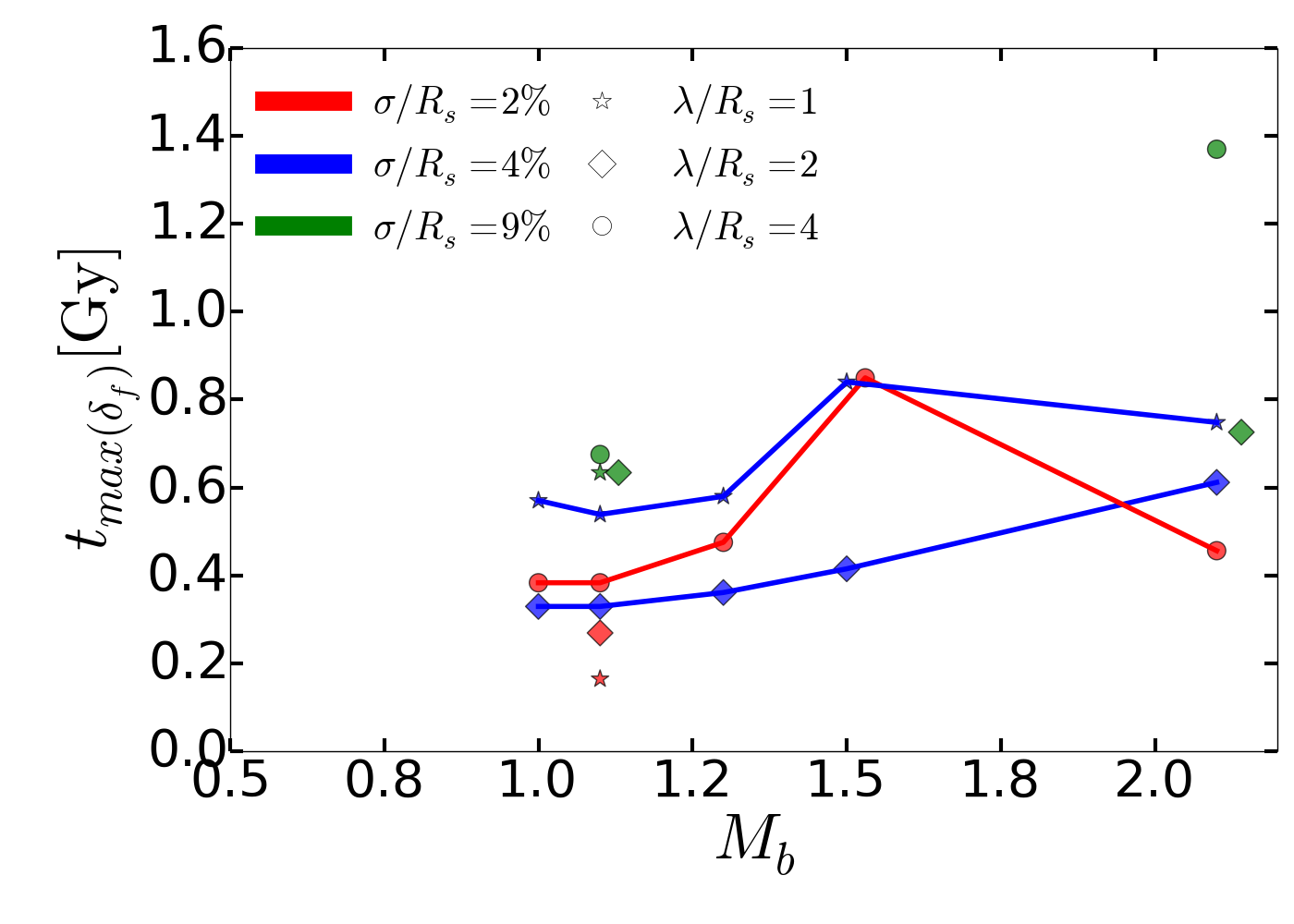}
 \caption{The time at which the maximum density contrast occurs as a function of $M_b$. Only simulations with a maximum density contrast of at least three are considered - $\rm{max}(\delta_{\rm{f}}) \geq 3$. The time axis is normalised to $R_s=4$ kpc (see equation \ref{eq:transformation} for rescaling). }
 \label{fig:varyMb_TimeOfmaxDensity}
 \end{figure}

In Fig. \ref{fig:2RsvaryMb184__maxDensityWithTime} we plot $\rm{max}(\delta_{\rm{f}})$ vs time for the same simulations as above, where $\rm{max}(\delta_{\rm{f}})$ is taken at each time step.
Simulations with $M_b=0.5$ do not reach high densities ($\delta_{\rm{f}} >3$) for the entire duration of the simulations for all tested initial conditions. 
However, all simulations with $M_b \geq 1.3$ show density contrasts of at least 6 independently of initial condition. With individual exceptions, these high density regions are mostly relatively short-lived and vanish after $\sim$ 300 Myr. We notice that the higher the $M_b$ the later the high densities seem to develop, see also Fig. \ref{fig:varyMb_TimeOfmaxDensity}. However, there are also exceptions to this behaviour. Moreover, there are systematic differences between the sets of simulations, which will be discussed later.

\begin{figure} 
 \includegraphics[width=\columnwidth]{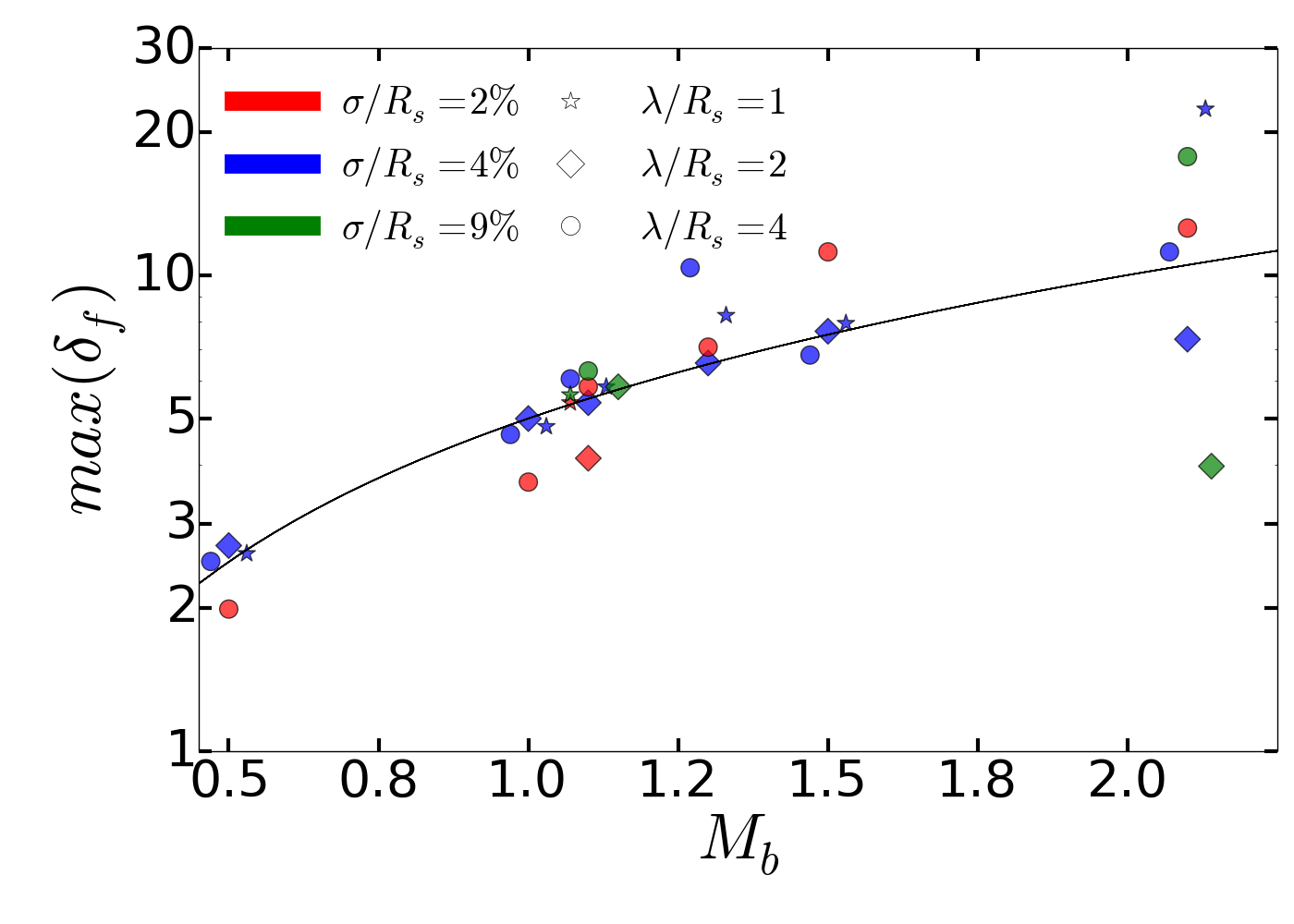}
 \caption{Maximum density contrast achieved during the simulation as a function of $M_b$ for various interface thicknesses (colors) and perturbation wavelengths (symbols). The black solid line shows the linear function $\delta_{\rm{f}}=5\times M_b$. }
 \label{fig:varyMb_maxDensity}
 \end{figure}

In Fig. \ref{fig:varyMb_maxDensity} we present the maximum density contrast as a function of $M_b$, where each symbol and color correspond to one set of parameters. 
In particular, the color indicates the interface thickness, $\sigma / R_s$, and the shape represents the wavelength $\lambda / R_s$. 
There seems to be a clear trend with increasing $M_b$ up to $M_b=1.5$, independently of the other parameters.
For larger $M_b$, there is instead a large spread of values for $\rm{max}(\delta_{\rm{f}})$ with no clear correlation with any values of $\lambda / R_s$ and $\sigma / R_s$. In Section \ref{sec_discussion} we discuss the possible origin of this trend or lack thereof for $M_b>1.5$.

\begin{figure} 
 \includegraphics[width=\columnwidth]{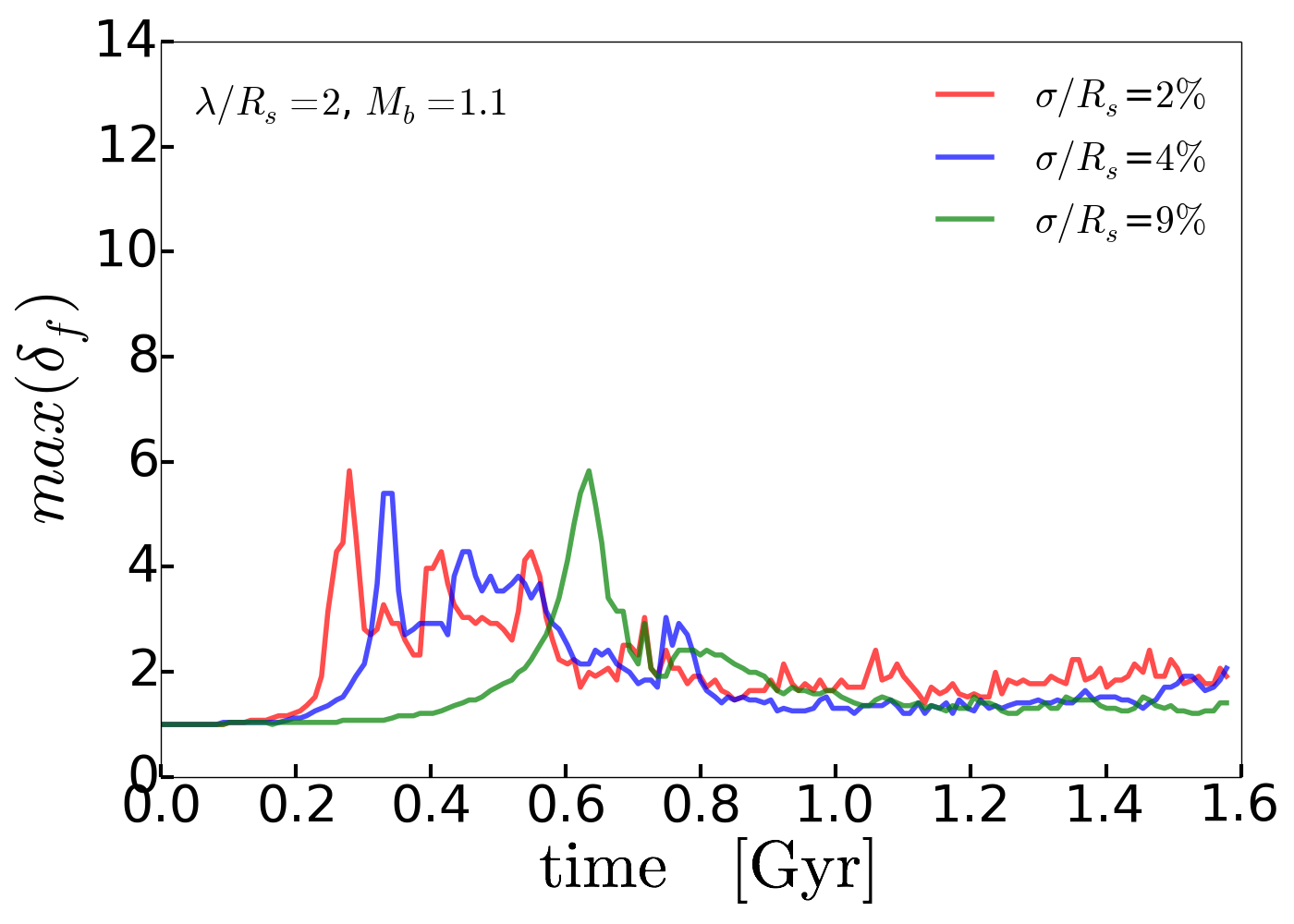}
 \caption{ Maximum density contrast as a function of time for three different interface thicknesses $\sigma / R_s$. All other parameters, e.g. the speed and wavelength are kept fixed. To avoid overlaps of the symbols we have slightly displaced them horizontally. The time axis is normalised to $R_s=4$ kpc (see equation \ref{eq:transformation} for rescaling).}
 \label{fig:varySigmaLambda4Rs_maxDensityWithTime}
 \end{figure}

\begin{figure} 
 \includegraphics[width=\columnwidth]{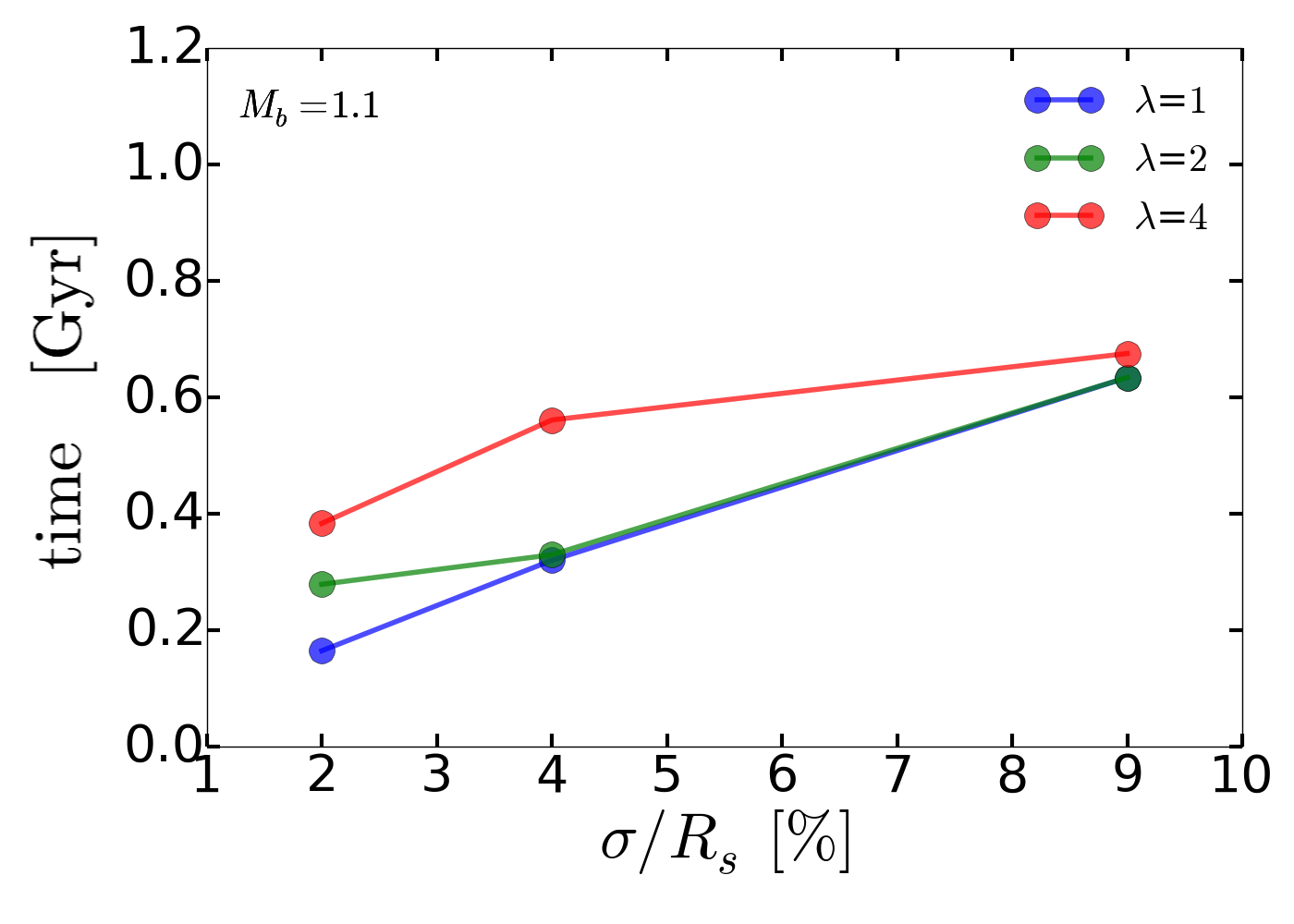}
 \caption{ The time at which the maximum density contrast occurs as a function of $\sigma / R_s$ for a fixed $M_b=1.1$ and for various perturbation wavelengths $\lambda / R_s$.}
 \label{fig:varySigma__maxDensity}
 \end{figure}
 
 \subsubsection*{When and where the high densities develop}
 
In Fig. \ref{fig:varyMb_TimeOfmaxDensity} we show at which time
the maximum density contrast $t_{\rm{max}(\delta_{\rm{f}})}$
occurs as a function of $M_b$ for a large set of parameters. 
In particular, we have only considered simulations that produced significant over-densities  
($\rm{max}(\delta_{\rm{f}})\geq3$) at any time, i.e. we have excluded models with $M_b=0.5$.
Symbols and colors represent different combinations of parameters as in Fig. \ref{fig:varyMb_maxDensity}.
Note that to avoid the overlap of the symbols we have slightly shifted them horizontally.  
To help the reader, we have joined with solid lines the symbols that correspond to the same set 
of $\sigma / R_s$ and $\lambda / R_s$, if simulation results are available for the entire range 
of $M_b$.
As is clear from the figure, the maximum over-densities develop faster for $1.0 \leq M_b \leq 1.3$. Once again,
there is a larger spread in $t_{\rm{max}(\delta_{\rm{f}})}$ for $M_b = 2.1$.

Now, we investigate the effect of the interface thickness $\sigma$ and $\lambda$ on $t_{\rm{max}(\delta_{\rm{f}})}$
fixing $M_b=1.1$. This is the typical $M_b$ value within the range where over-densities develop faster,
as shown in Fig. \ref{fig:varyMb_TimeOfmaxDensity}.
In particular, Fig. \ref{fig:varySigmaLambda4Rs_maxDensityWithTime} shows how $\rm{max}(\delta_{\rm{f}})$ changes with time for different 
values of $\sigma$ for a constant $\lambda / R_s = 2$ and $M_b=1.1$.
Similarly, in Fig. \ref{fig:varySigma__maxDensity}, we show how $t_{\rm{max}(\delta_{\rm{f}})}$ changes with $\sigma$ for three different
values of $\lambda$ and $M_b=1.1$. As shown in Fig. \ref{fig:varySigmaLambda4Rs_maxDensityWithTime} and \ref{fig:varySigma__maxDensity}, the thinner the interface between hot halo and filament, the shorter it takes for high densities to develop. 
The same trend also applies to $M_b=2.1$ (not shown here).

Through a visual inspection of the simulation outputs we have noticed that high densities always start developing at the interface. 
Afterwards, the majority of these ``travel" to the center of the filament and they become less dense. 
Once they reach the center of the filament, the densities increase again due to the interaction with 
overdensities coming from the opposite side.
For $M_b \geq 2.1$ high densities do not always penetrate within the filament and disappear after
some time. 

\begin{figure} 
 \includegraphics[width=\columnwidth]{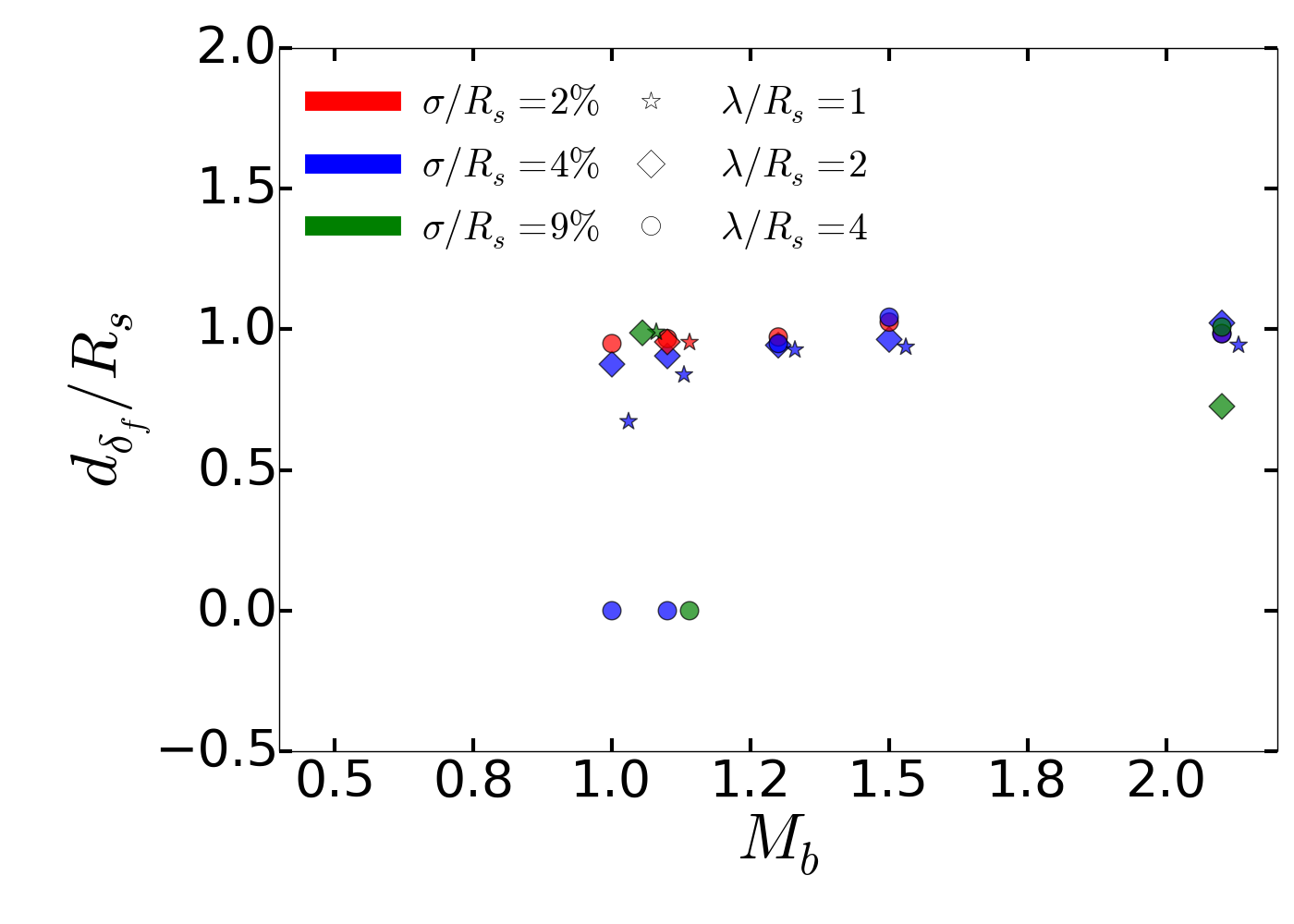}
 \caption{Position within the filament where the maximum density occurs. $x=0.0$ corresponds to the filament center, $x=1.0$ to the interface.}
 \label{fig:locationMaxDensityMb}
 \end{figure}

In order to quantify this behaviour, we show in Fig. \ref{fig:locationMaxDensityMb} the distance between the position of $\rm{max}(\delta_{\rm{f}})$ and the center of the filament
in function of $M_b$ for different values of $\sigma$ and $\lambda$.
As is clear from the figure, $\rm{max}(\delta_{\rm{f}})$ is either located at the interface or at the center of the filament. 
In particular, only perturbations with a large wavelength $\lambda/R_s = 4$ develop the $\rm{max}(\delta_{\rm{f}})$ at the center.
For $M_b\geq1.3$ all $\rm{max}(\delta_{\rm{f}})$ are found at the interface. Conversely, simulations
with lower $M_b$ produce high densities both at the center or at the interface depending on the value of
$\lambda$ and $\sigma$. 
Because we are only tracing the maximum densities, 
this result does not imply that there are no overdensities in the rest of the filament.  
We notice that in most cases, there are high densities ($\delta_{\rm{f}} \geq 3$) developing at the center even when 
$\rm{max}(\delta_{\rm{f}})$ is occurring at the interface, see as an example Fig. \ref{fig:Mb1.3-extreme} (middle).

\subsubsection*{Morphology of the high-density regions}

\begin{figure} 
 \includegraphics[width=\columnwidth]{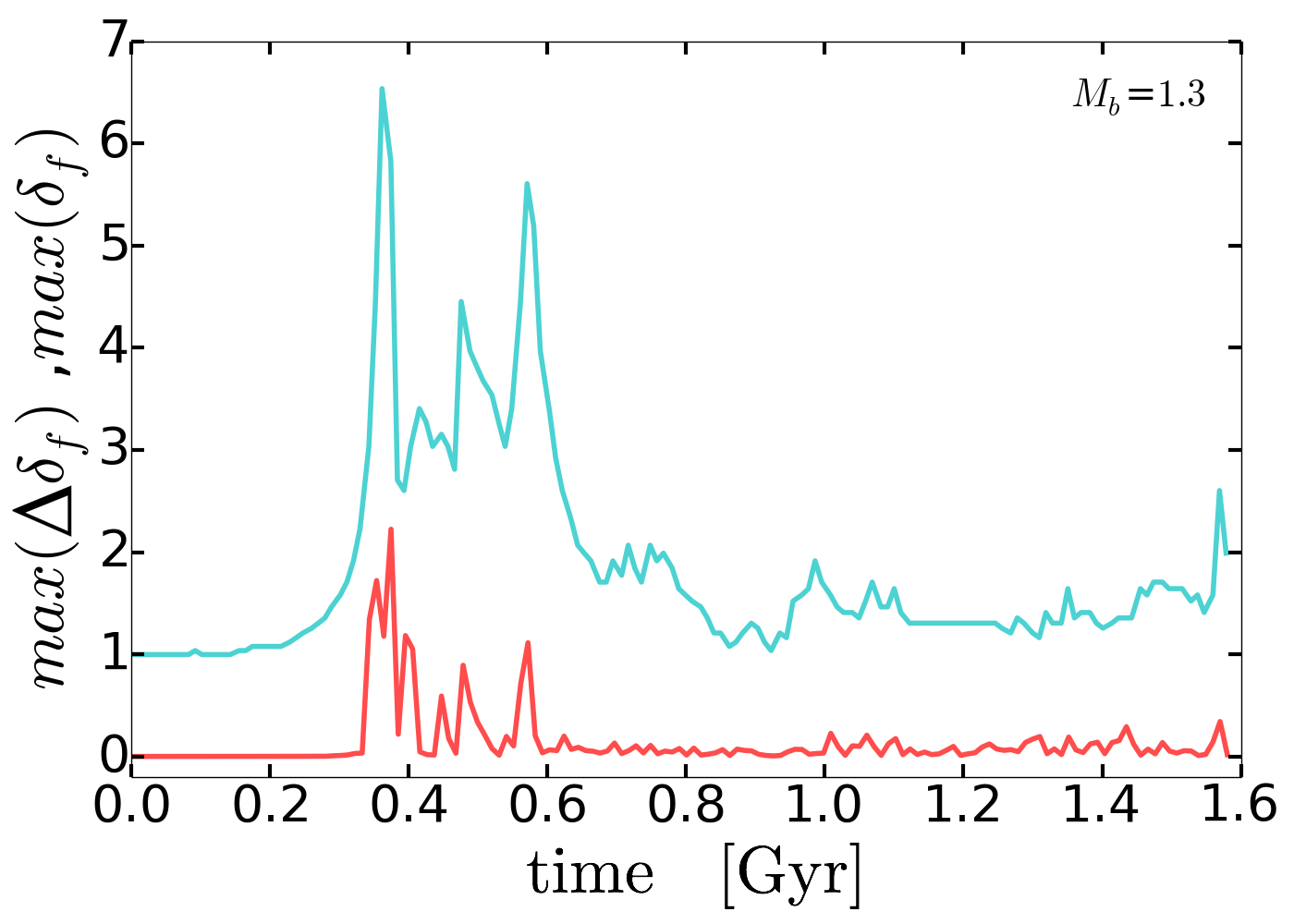}
 \caption{ (red): Maximum density difference between the cell with the maximum density contrast and all its neighbouring cells vs time. (cyan): Max density contrast vs time, as above. $M_b=1.3$, $\lambda / R_s=2$ and $\sigma / R_s = 4\%$ for this simulation.  }
 \label{fig:densityGradient}
 \end{figure}

In this Section we present the results of one simulation with $M_b=1.3$, $\lambda / R_s=2$ and $\sigma / R_s = 4\%$ as a representative case
for the development of high densities regions forming in the power law tail of the PDF. 
We use the maximum difference between the density of the cell corresponding to $\rm{max}(\delta_{\rm{f}})$ and the densities of its 8 neighbours (indicated as $\rm{max}(\Delta \delta_{\rm{f}})$ in the remainder of this Section). 
In Fig. \ref{fig:densityGradient} we show both $\rm{max}(\Delta \delta_{\rm{f}})$ and $\rm{max}(\delta_{\rm{f}})$ as a function of time as the red and cyan solid line, respectively. 
We find that $\Delta (\delta_{\rm{f}}) << (\delta_{\rm{f}})$ at any time. This suggests the seeds developing in our simulations are real and larger than a few simulation cells.
Clearly, the time evolution of $\rm{max}(\Delta \delta_{\rm{f}})$ traces the corresponding evolution of $\rm{max}(\delta_{\rm{f}})$ very well.

\begin{figure} 
 \includegraphics[width=\columnwidth]{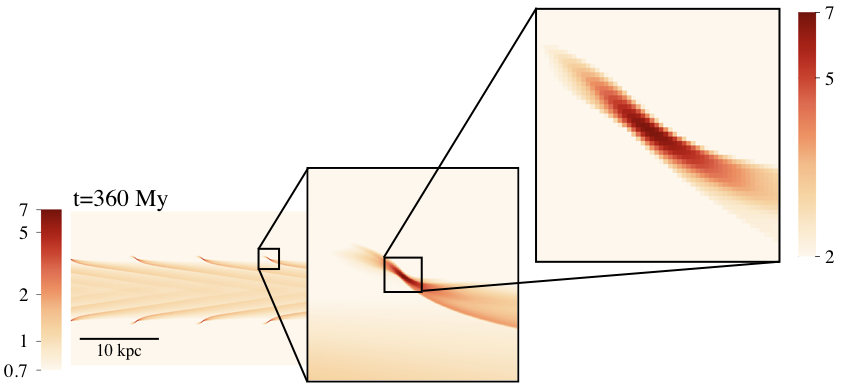}
 \caption{ Zoom in on an overdense region developing at $t=360$ Myr for $M_b=1.3$, $\lambda / R_s=2$ and $\sigma / R_s = 4\%$.}
 \label{fig:seedSnapshot}
 \end{figure}

The resolved nature of the the densest regions made up of numerous pixels can also be verified by a direct visual inspection of the simulation output 
at the moment in which $\rm{max}(\delta_{\rm{f}})$ is reached, as shown in Fig. \ref{fig:seedSnapshot} (see also Fig. \ref{fig:Mb1.3-extreme}).
We also note that their morphology appears elongated along the direction of the filament flow. We will discuss the possible origin of this morphology in Section \ref{sec_limitations}.

 \subsection{Filament disruption and accretion rates}

\subsubsection*{How much cold gas from filament accretion reaches the galaxy?}
Now, we investigate how much cold gas reaches the galaxy at the center of the halo. 
To answer this, we investigate the net cold gas mass flux ($j_{\rm{m}}$), defined as
 \begin{equation}
 j_{\rm{m}}= \int \rho_{\rm{cold}} v_x dy\ ,
 \end{equation}
as a function of time
using our fiducial suite of simulations with $\lambda / R_s = 2$,  $\sigma / R_s=4 \%$.

 \begin{figure} 
 \includegraphics[width=\columnwidth]{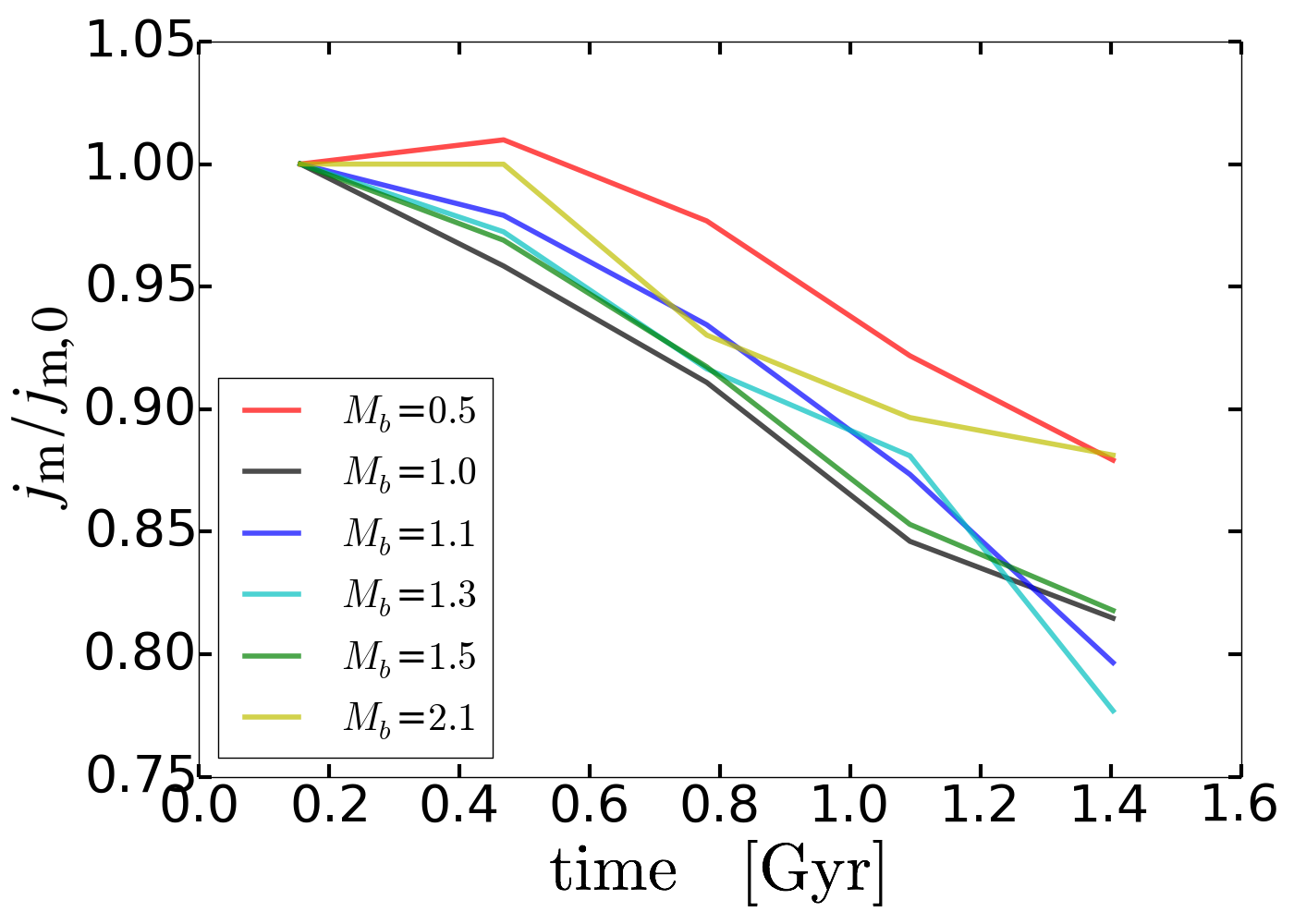}
 \caption{ Normalised net mass flux of the cold gas through a vertical slice of the simulation box as a function of time and $M_b$ at fixed $\lambda / R_s = 2$,  $\sigma / R_s=4 \%$. }
 \label{fig:massFlux}
 \end{figure}

In Fig. \ref{fig:massFlux}, we show the net mass flux of the cold gas ($T<5\times10^4$ K) 
through a vertical slice of the simulation box as a function of time normalised to the initial value.
Because we are using only one slice, the net flux varied strongly over short time scales. In order to show only the average
behaviour on large time scales we have binned our results over 0.3 Gyr in time.
As it is clear from the figure, the mass flux of cold gas declines monotonically for all tested cases. 
Filaments that are nearly sonic show the largest decrease in mass flux by 20 percent.

We have seen in Section \ref{sec:devHighDensity} that filaments with $M_b$ in the range $1.0-1.3$
produce largest densities faster, possibly indicating a connection between the development of high densities and a reduction in accretion rate.

\subsubsection*{Do filaments disrupt during accretion?}

If accreting filaments were to disrupt, this would likely have an impact on their morphology, which could be observed. 
In order to study the filament disruption we introduce a quantity called Perturbation Depth (PD)
defined as:

\begin{equation}
 \begin{aligned}
& PD=\rm{max}( | y_{\rm{interface}}-y_{\rho \leq \rho_{\rm{thresh}} } | ) \\
 & \rho_{\rm{thresh}}= \frac{1}{3} \rho_{\rm{fil}} \\
 \end{aligned}
 \end{equation} 
 
 where $y_{\rm{interface}}$ is the initial y-location of the interface (as defined in the initial conditions)
 and $y_{\rho \leq \rho_{\rm{thresh}}}$ are the y-locations of simulation cells with 
 gas densities below $\rho_{\rm{thresh}}$.
In particular, the PD quantifies how deep the surrounding halo gas with a density smaller than a certain 
threshold penetrates into the filament. This approach is different from \cite{Padnos2018} where tracers in the fluid are used to determine whether the filament has fully disrupted.
We tested different values for this threshold and verified that 
a value of $1/3 \rho_{\rm{fil}}$ best captures 
how far the surrounding halo gas penetrates into the filament
without contamination from low density regions in the center. 
In order to be less sensitive to numerical noise,
we re-grid our simulation box into cells with sizes
that are nine times larger than the initial ones.

\begin{figure} 
 \includegraphics[width=\columnwidth]{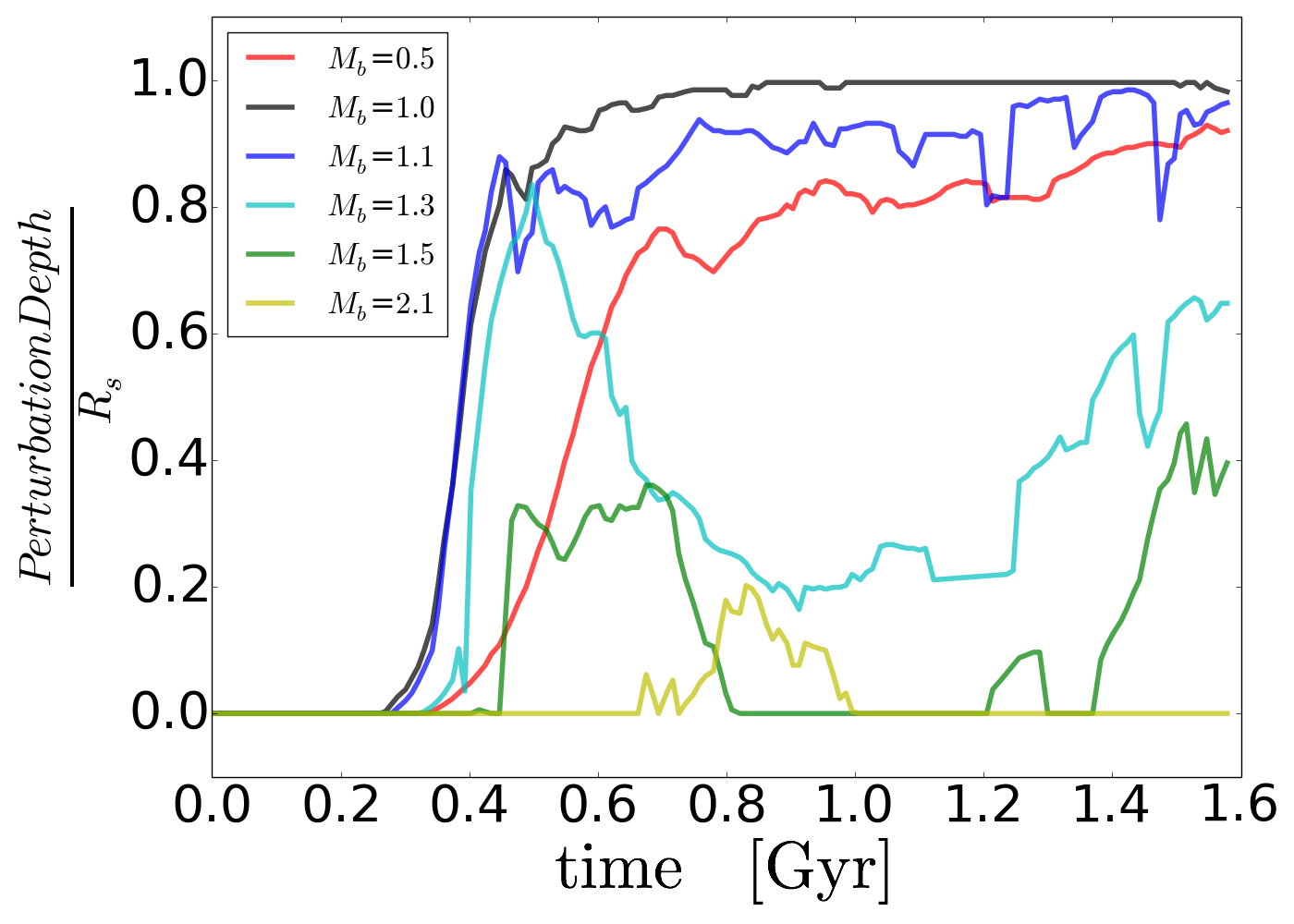}
 \caption{ Filament disruption as measured through the Perturbation Depth normalized to the filament radius $R_s$ as a function of time. The time axis is normalised to $R_s=4$ kpc (see equation \ref{eq:transformation} for rescaling). The other properties used in the simulations are $\lambda/R_s=2$ and $\sigma=4 \%$.}
 \label{fig:2RsvaryMb184__PerturbationDepth}
 \end{figure}

In Fig. \ref{fig:2RsvaryMb184__PerturbationDepth}, we show how the normalised PD evolves with time for different $M_b$
and fixing $\lambda/R_s=2$ and $\sigma/R_s= 4 \%$.
All simulations with $M_b \leq 1.3$ reach a PD of 0.8$R_s$.

 \begin{figure} 
 \includegraphics[width=\columnwidth]{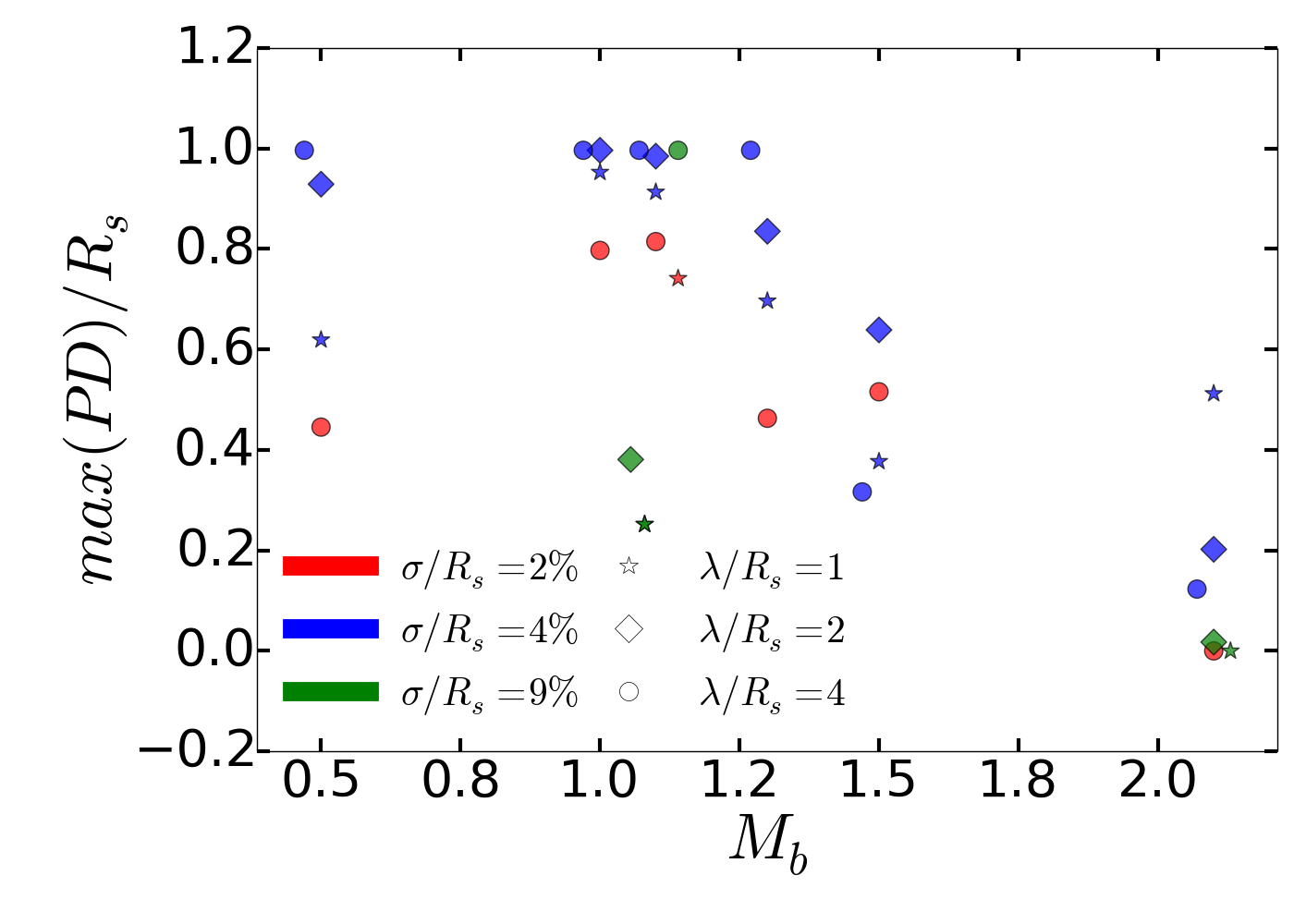}
 \caption{The maximum Perturbation Depth reached throughout the duration of the simulation as a function of $M_b$. }
 \label{fig:varyMb_MaxPerturbationDepth}
 \end{figure}
 
 In Fig. \ref{fig:varyMb_MaxPerturbationDepth}, we show the maximum Perturbation Depth, $\rm{max}(PD)$ reached throughout 
 the duration of the simulation as a function of $M_b$ for a large range of parameters. 
 Simulations with $M_b \geq 1.3$ show a small Perturbation Depth ($PD< 0.6 R_s$) for all explored parameters. 
 This result, together with the time evolution of $j_{\rm{m}}$ presented in \ref{fig:massFlux}, suggests that filaments 
 with $M_b \geq 1.3$ are less likely to be disrupted. 
 We also remark that there is a preferential interface thickness, $\sigma / R_s = 4 \%$ (blue symbols in Fig. \ref{fig:varyMb_MaxPerturbationDepth}) 
 where filaments tend to have a larger $\rm{max}(PD)$, e.g. they may disrupt more easily.

\section{Discussion}
\label{sec_discussion}

In the previous section we have found that, when properly resolved,
accreting filaments into hot haloes develop instabilities
that result in:
i) broad ``cold gas" density distributions that are qualitatively consistent
with skewed log-normal Probability Distribution Functions
ii) additional components to the skewed log-normal PDF in form of high and low density tails for $M_b>1.3$ (where $M_b$ indicates the filament velocity with respect to the hot halo sound speed; see Fig. \ref{fig:MbPDFs});
iii) maximum density contrasts up to a factor of 20, depending on the value of $M_b$ (see Fig. \ref{fig:varyMb_maxDensity});
iv) faster development of high density regions 
for $1 \leq M_b \leq 1.3$ and thinner interface regions (see Fig. \ref{fig:varyMb_TimeOfmaxDensity} \& \ref{fig:varySigmaLambda4Rs_maxDensityWithTime});
v) more likely filament disruption and possibly smaller accretion rates onto galaxies 
for $1<M_b<1.5$ (see Fig. \ref{fig:varyMb_MaxPerturbationDepth}).
These results suggest that a $M_b \geq 1$ is a promising condition for the formation
of high density structures within haloes due to filament accretion and a moderate reduction in accretion rate. 
In this section, we discuss the possible origins and the implications of these results.

\subsection{Origin of the broad gas density distribution and fate of the filament}

There is a large literature on the origin of log-normal gas density distribution through turbulent processes, traditionally applied to the Interstellar Medium (ISM)
\citep{Padoan1997, Vazquez1994, Passot1998, Padoan2002}. In particular, previous studies suggest
that a succession of compressive or expansive waves can interact with each other 
producing density fluctuations on different scales (that are uncorrelated 
and continuous) which would result in log-normal distributions \citep[e.g.][]{Passot1998,Squire2017}. 
However, recent studies have found deviations from an exactly log-normal behaviour and argued that a 
skewed log-normal gas density distribution better reproduces the simulation results \citep{Federrath2010, Federrath2008, Hopkins2013b, Molina2012, Nolan2015}.
In particular, these authors argued that a log-normal density distribution on multiple scales would indeed violate mass conservation
and that deviation from this shape are therefore to be expected \citep[e.g.][]{Hopkins2013b}.
Moreover, \cite{Nolan2015} have found a relation between the sonic Mach number ($M_s=v_s/c_s$)
and the variance (S) of the skewed log-normal gas density distribution, in the case of non-isothermal adiabatic turbulence:

\begin{equation} \label{eq:sigma_Ms}
\begin{aligned}
S^2 &= \rm{ln}(1+bM_s^{(5 \gamma + 1)/3}), \quad\quad &b M_s \leq 1 \\
S^2 &=  \rm{ln}\Bigg (1+\frac{(\gamma +1)b^2 M_s^2}{(\gamma +1)b^2 M_s^2+2} \Bigg ), \quad\quad & b M_s > 1
\end{aligned}
\end{equation}

where $\gamma$ is the adiabatic index and
$b$ is a constant for the turbulence driving parameter that depends on the mode mixture induced by the turbulent forcing mechanism
(e.g. curl-free or divergence-free). Although the precise value of $b$ is unclear when modes are mixed, 
several studies suggest that $b$ ranges between 1/3 and 1 \citep[see e.g. ][for further details]{Federrath2008}. 

Although our simulations are different from isotropic turbulent studies, we will use
these results as a possible ansatz for the origin of the skewed log-normal 
gas density distribution in our case. 
Instead of externally driving turbulence, our simulations can in principle develop turbulence more naturally due to KHI, i.e.,
through the disruption of vortices via secondary instabilities, vortex pairing and nonlinear interactions \citep{Lecoanet2016}.
Note that $M_s$ indicates the velocity of a gas parcel with respect to its own sound speed while
$M_b$, in our study, represents the filament velocity with respect to the sound speed of the static, hot halo gas.  

In order to determine whether turbulence could be the origin of a skewed log-normal PDF also for accreting filaments, 
we test whether the relation \ref{eq:sigma_Ms} holds in our case. To this end we need to compute the value of $M_s$. 
In particular, we examined a simulation with  $M_b=1.3$ at a given time (see middle panel of Fig. \ref{fig:Mb1.3-extreme}) for which
the PDF is well approximated by a skewed log-normal, finding values for $M_s$ for cells with temperatures below $5\times10^4$ K. 
In order to derive these values, we have first computed the velocity dispersion $v_s$ for each cell with temperatures below $5\times10^4$ K including all neighbouring cells 
across a set of of scales  ($L_c$). These scales range from the thickness of the interface (20 cells) to half the filament radius (340 cells).
We have then divided $v_s$ of each cell by its own sound speed and obtained in this way a range of $M_s$ that spans from 0.02 to 2.1
for all "cold" cells and scales. Note that, as expected, $M_s$ increases with $L_c$.
What are the expected values of S in this case using equation \ref{eq:sigma_Ms}?
Using the range of $M_s$ above, we obtain S$=[0.002, 0.3]$ (S$=[0.0013, 0.2]$) for $b=1$ ($b=1/3$).
This range of S is consistent with the measured dispersion of the skewed log-normal PDF in our simulation, namely $S=0.009$ (see dashed line in Fig. \ref{fig:Mb1.3-extreme}),
suggesting that this component of the PDF may originate from turbulence, at least for this particular simulation.

How do these results change considering simulations with different $M_b$, $\lambda / R_s$ and interface thickness $\sigma / R_s$?
As shown in Fig. \ref{fig:varyMb_maxDensity}, we do find a possible linear correlation between max$(\delta_{\rm{f}})$ and $M_b$, albeit with a large scatter at $M_b=2.1$,
from which one could also expect a linear relation between S and $M_b$, at least within a range of $M_b$. Moreover, 
this result seems fairly independent 
of the value of $\lambda / R_s$ and $\sigma / R_s$, at least in the tested range of values.
This suggests that the skewed log-normal PDF could originate from turbulence in the majority of our simulations. 

While the skewed-log-normal component is consistent with turbulence as discussed above, the high-density power-law tail cannot be easily explained in this scenario. What is the origin of this component? 
Previous studies have suggested that KHI are unable to grow at the surface of the filament for $M_b>M_{\rm{crit}}=(1+\delta^{-1/3})^{3/2}$ \citep{Mandelker2016}. 
Note that $M_{\rm{crit}}$ is exactly 1.3 for our simulations, for which $\delta=152$ and this is the same $M_b$ above which we notice the appearance of high density tails in the PDFs (see Fig. \ref{fig:MbPDFs}). 
Although KHI are suppressed at the filament interface above $M_{\rm{crit}}$, we stil expect that 
instabilities could be present as a consequence of reflected sound waves \citep[also called ``body modes", see][]{Mandelker2016}.
Such sound waves reflect off the oppositely located interfaces of the filament and eventually compress the gas at the filament surface.
Alternatively, ``body modes" may slightly alter the overall shape of the gas stream exposing small regions at the interface with the hot gas to a large velocity gradient resulting in extra compression therein.
In both cases, these processes could lead to the production of small "chains" of strongly compressed and therefore very dense cold gas, as observed in our
simulations at $M_b\geq1.3$ (see Figs. \ref{fig:Mb1.3-extreme}, bottom left \& \ref{fig:seedSnapshot}).
In particular, the individual dense gas clumps have the highest density at their center, 
which decreases with distance, resulting in a continuous density distribution that could resemble a power law. 

We note that our fast flowing filaments never develop the highest densities at their center (Fig. \ref{fig:locationMaxDensityMb}) and they never disrupt, at least for the range of explored parameters. 
This is consistent with the picture discussed above, because ``body modes" can destroy the filament only if the perturbation length ($\lambda$) 
is larger than  $10R_s = \lambda_{\rm{crit}}$ \citep{Padnos2018}, which we did not consider here. 
On the other hand, filaments with  $M_b<1.3$, that have sufficiently large perturbation wavelength ($\lambda > 2R_s$) and with an interface thickness of $\sigma/R_s=4\%$ are most likely to disrupt. 
Are these filaments able to reach the central galaxies before they are disrupted? 
This depends on the halo crossing time, $t_{\rm{cross}}$ that in our simulations ranges between 0.3 Gyr and 1.3 Gyr for $M_b=2.1$, 
and 0.5 respectively for a halo mass of $M_{\rm{vir}}=10^{12.5}$\(\textrm{M}_\odot\) at redshift $z \sim 3$.
Fig. \ref{fig:2RsvaryMb184__PerturbationDepth} shows that, if a filament with radius of 4 kpc is subject to disruption, the disruption takes place at $t_{\rm{disrupt}}\sim0.3 - 0.4 \;\rm{Gyr} < t_{\rm{cross}}$, for this specific set of simulations. This is approximately half the filament sound crossing time, which is $t_{\rm{sound}} \sim 0.6 \;\textrm{Gyr}$ for a filament with $R_s =4$ kpc and $M_b \leq 1.3 $. 
These results are consistent with previous findings \citep[e.g.][]{Padnos2018}. 
We remark that our choice of $M_b$ is motivated by cosmological simulations, which suggest that infalling filaments will have a constant velocity of about the halo virial velocity \citep{Dekel2009}. Depending on assumptions of the virial temperature, the halo mass and the velocity of the filament, various Mach numbers are plausible, ranging from $M_b = 0.5$ up till 5.0.

Lastly, we comment on possible implications of our results in context of recent observations of Ly$\alpha$ nebulae around quasars that suggest a multiphase, possibly extremely clumpy intergalactic and circumgalactic medium. A possible ansatz for this density distribution is a log-normal PDF of the emitting gas of the Nebulae with dispersion $\sigma_s \sim 2$ defined as in eq. \ref{eq:PDF_lognormal} \citep{Cantalupo2018}. 
Although our PDFs have a much smaller dispersion than by these observations,
they may be broad enough to trigger a succession of processes which may lead to densities in line with these recent findings. These processes and their effect on the density distribution of the gas will be discussed in the next section.

\subsection{Current limitations \& future prospects}
\label{sec_limitations}

In this section, we discuss possible processes, not yet included in our idealised experiment, e.g. radiative cooling and gravity, or numerical limitations which could lead to a broadening or narrowing of the gas density distribution or a modification of our results.

In particular, we expect that when high densities are formed (either due to KHI or "body modes" as discussed above), radiative cooling could further enhance them which will then
possibly lead to a broadening of the gas density distribution and also extend the cold clumps ``lifetime". 
Furthermore, parts of the filament in which the densities have been enhanced by KHI may be subject to gravitational instabilities, especially when this effect is combined with radiative cooling, leading to a more pronounced broadening of the gas density distribution. 

On the other hand, cooling and gravity could also prevent KHI to develop in the first place, at least at the interface, through stabilisation effects \citep[e.g.][]{Vietri1997}. 
Similarly, KHI may be prevented by thermal conduction through smoothing of small scale inhomogeneities and velocity gradients \citep{Armillotta2016}.
However, if KHI grow and ``eddies" develop, thermal conduction may avoid subsequent KHI within these regions avoiding their fragmentation and mixing with the hot material. 
 Note, that since we are using a grid code, there is already numerical conduction in our simulation, as determined by the cell size.  
 Lastly, numerous studies have shown that magnetic fields have a stabilising effect on KHI \citep[][ and references therein]{Vietri1997}, however there are little observational constraints on the strength of magnetic fields in the
 CGM of galaxies at $z \sim 3$. 
 
What is the effect of numerical resolution on our results? Because of limited computational time, it is difficult to
assess the numerical convergence of our results by increasing the simulation resolution. However, we have
re-run some simulations at lower resolution as shown in the Appendix in order to have at least a qualitative
assessment of resolution effects. In particular, we have tested resolution effects
on the value of max($\delta_{\rm{f}}$), which could be expected to be particularly prone to changes in resolution, finding very limited changes. 
 
Although performing simulations in 3D might change the results presented here \citep[see also][]{Mandelker2018}, we think that including radiative cooling could be a more important step towards 
a better characterisation of the real CGM density distributions. We plan therefore to implement radiative cooling in our next set of simulations for future studies.

\section{Summary \& Conclusions}
\label{sec_summary}
Our standard cosmological model predicts that cosmic filaments
should play a crucial role in galaxy evolution transporting
gas from the intergalactic medium into galaxies.
In this work, we have explored how hydrodynamical effects
such as Kelvin-Helmoltz instabilities (KHI) affect the physical properties of
the accreting streams within the circumgalactic medium (CGM) of high-redshift galaxies
with a particular focus on the formation and evolution of high density structures.
The presence of broad gas density distribution in the cold, Ly$\alpha$ emitting
phase of the CGM has been suggested by recent observations of extended
emission around high-redshift quasars \citep[e.g.][]{Cantalupo2014, Borisova2016, Cantalupo2018}.

In order to achieve the required spatial resolution to study in detail the formation
and evolution of KHI, we have performed idealised simulations with the hydrodynamical code 
RAMSES \citep{ramses} and explored a large parameter space in filament and perturbation properties, such as filament Mach number, initial perturbation wavelength, and thickness of the interface between the filament and the halo.
We find that the time averaged density distribution of the cold gas is qualitatively consistent with a skewed log-normal probability 
distribution function (PDF) plus an additional component in form of a high density tail for simulations with $M_b>1.3$.
We explored the possible origin of the skewed log-normal PDF and found that the broadness of the PDF is related
to the value of $M_b$ as qualitatively expected in a turbulence-driven scenario in which the turbulence is produced by the 
nonlinear development of KHI.
We speculate that the high-density tail of the PDF could be produced by the interaction between the hot gas and coherent 
density waves produced within the filament by the development of reflecting sound waves at high $M_b$.

In our simulations the cold gas densities can reach values as high as 20 times the initial filament densities but only
for a very limited time before being dissolved. In all explored cases the simulated cold gas density distributions are
much narrower than the one implied by the observations, suggesting that additional physical mechanism or processes
are needed. However, our simulations suggest that cosmological accretion, in some particular cases, 
could be a viable mechanisms to produce the `seeds' for the possible growth of higher over-densities 
in the CGM of high-redshift galaxies.

\section*{Acknowledgements}
We thank Romain Teyssier for making his code, RAMSES publicly available and gratefully acknowledge support from the Swiss National Foundation grant PP00P2\_163824. Also, I want to thank my husband for his consistent and ongoing support.
\addcontentsline{toc}{section}{Acknowledgements}




\bibliographystyle{mnras}
\bibliography{refs} 





\appendix 
\section{Resolution Study}
 
\begin{figure}
 \includegraphics[width=\columnwidth]{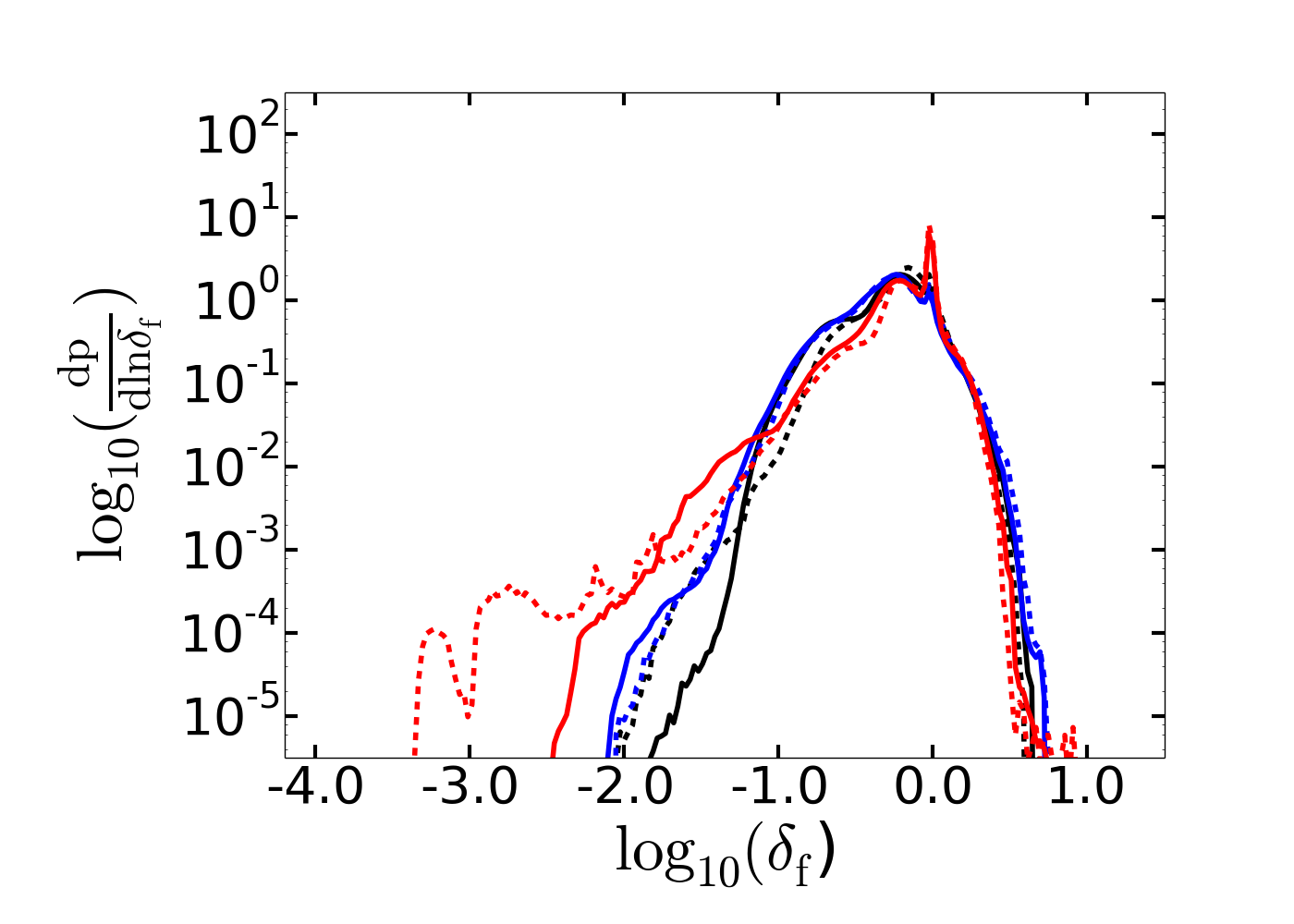}
 \includegraphics[width=\columnwidth]{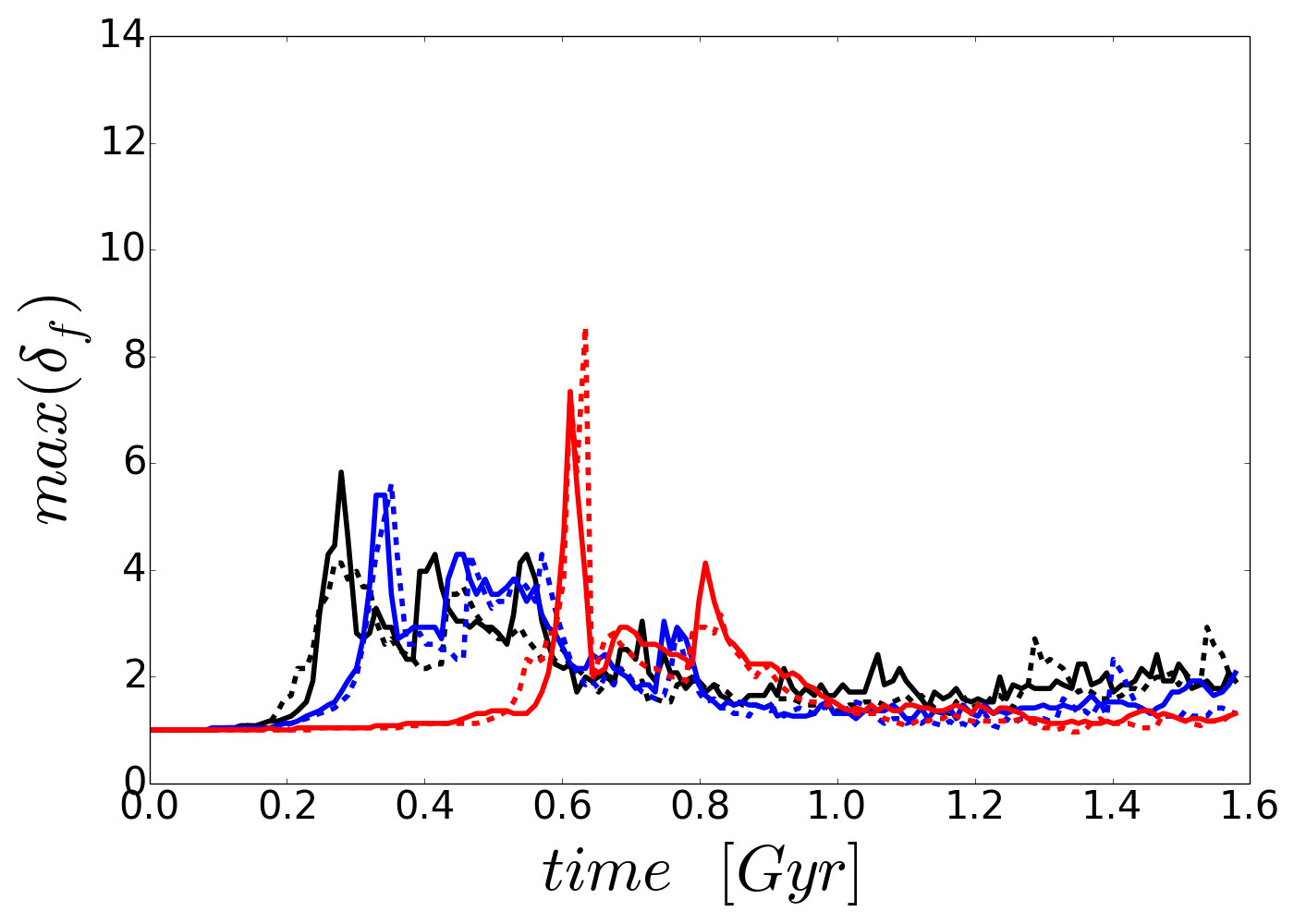}
 \caption{Time averaged density distribution (top) and time evolution of maximum density contrast (bottom) at high resolution ($2^{12}$ cells by side, solid lines) and low resolution ($2^{11}$ cells, dashed lines). Parameters used are as follows: $M_b=1.1$, $\lambda/R_s=2$, $\sigma/R_s=2\%$ (black); $M_b=1.1$, $\lambda/R_s=2$, $\sigma/R_s=4\%$ (blue); $M_b=2.1$, $\lambda/R_s=2$, $\sigma/R_s=4\%$ (red).
 }
 \label{fig:Appendix:resolution}
 \end{figure}

We have rerun some simulations with lower resolution in order to get an estimate of how resolution may be affecting our results. Fig. \ref{fig:Appendix:resolution} (top) shows the evolution of max($\delta_{\rm{f}}$) with time for two resolutions. Although max($\delta_{\rm{f}}$) may be prone to numerical noise, we find the effect of changing resolution to be small. In in Fig. \ref{fig:Appendix:resolution} (bottom) we show the PDFs for two different resolutions and likewise, we find the difference of the PDFs to be too small to be significant. 

We do not find any significant or systematic dependence on resolution, especially at high densities.
 
\bsp	
\label{lastpage}
\end{document}